\pdfoutput=1
\documentclass[10pt,sigconf]{acmart}
\usepackage{comment}
\usepackage{subfigure}
\usepackage{multirow}
\usepackage{pifont}
\usepackage{balance}
\usepackage{textcomp}
\usepackage{booktabs}

\newcommand{\sect}[1]{\textsection#1}
\newcommand{\rev}{}

\begin{document}

\copyrightyear{2019} 
\acmYear{2019} 
\setcopyright{acmcopyright}
\acmConference[MobiCom '19]{The 25th Annual International Conference on Mobile Computing and Networking}{October 21--25, 2019}{Los Cabos, Mexico}
\acmBooktitle{The 25th Annual International Conference on Mobile Computing and Networking (MobiCom '19), October 21--25, 2019, Los Cabos, Mexico}
\acmPrice{15.00}
\acmDOI{10.1145/3300061.3300118}
\acmISBN{978-1-4503-6169-9/19/10}
\settopmatter{printacmref=true}
\fancyhead{}
\title[Touch-to-Access Device Authentication Using Induced BEPs]{Towards Touch-to-Access Device Authentication Using Induced Body Electric Potentials}

\author{Zhenyu Yan}
\affiliation{\institution{Nanyang Technological University}
}
\email{zyan006@ntu.edu.sg}

\author{Qun Song}
\affiliation{\institution{Nanyang Technological University}
}
\email{song0167@ntu.edu.sg}

\author{Rui Tan}
\affiliation{\institution{Nanyang Technological University}
}
\email{tanrui@ntu.edu.sg}

\author{Yang Li}
\authornote{This work was completed while Yang Li was with Advanced Digital Sciences Center, Illinois at Singapore.}
\affiliation{\institution{Shenzhen University}
}
\email{yli@szu.edu.cn}

\author{Adams Wai Kin Kong}
\affiliation{\institution{Nanyang Technological University}
}
\email{adamskong@ntu.edu.sg}

\begin{abstract}
  This paper presents TouchAuth, a new touch-to-access device authentication approach using induced body electric potentials (iBEPs) caused by the indoor ambient electric field that is mainly emitted from the building's electrical cabling. The design of TouchAuth is based on the electrostatics of iBEP generation and a resulting property, i.e., the iBEPs at two close locations on the same human body are similar, whereas those from different human bodies are distinct. Extensive experiments verify the above property and show that TouchAuth achieves high-profile receiver operating characteristics in implementing the touch-to-access policy. Our experiments also show that a range of possible interfering sources including appliances' electromagnetic emanations and noise injections into the power network do not affect the performance of TouchAuth. A key advantage of TouchAuth is that the iBEP sensing requires a simple analog-to-digital converter only, which is widely available on microcontrollers. Compared with existing approaches including intra-body communication and physiological sensing, TouchAuth is a low-cost, lightweight, and convenient approach for authorized users to access the smart objects found in indoor environments.
\end{abstract}



\maketitle

\section{Introduction}
\label{sec:intro}

The indoor environments are increasingly populated with smart objects. It is estimated that by 2022, a typical family home could contain more than 500 smart devices \cite{number-of-smart-objects}. Managing the access with many objects, including accessing the information on them or granting them to access certain information, becomes challenging. Typing password is tedious and infeasible for the objects without a keyboard or touchscreen. Biometrics-based user authentication suffers various shortcomings. Fingerprint scanning requires a well positioned finger press. Moreover, due to cost factor, small objects will unlikely have fingerprint scanners. Face recognition solutions require face positioning and are costly \cite{costly-faceid}. Voice recognition-based access can be disturbing in certain environments, e.g., an open-plan office with colleagues, a bedroom with sleeping buddies, etc. Moreover, defining a separate voice passphrase for each smart object to avoid incorrect invoking may result in too many passphrases.

In this paper, we aim to develop a low-cost and convenient {\em touch-to-access} scheme that can be easily implemented on smart objects found in indoor environments. Specifically, a simple touch on an object allows an authorized user to access the object.
This scheme will not require non-trivial interferences for user interactions, e.g., touchscreen. Different from integrating user identification (e.g., fingerprint scanning) into the objects, we resort to a {\em device authentication} approach that offloads the user's identity to a personal {\em wearable token} device (e.g., a smart watch or bracelet) and uses the token to access a touched object that has been previously paired with the token. This touch-to-access device authentication approach can greatly improve the user's convenience and experience in interacting with the smart objects. For instance, in a home with multiple residents, when a user wearing his token turns on a TV set using a smart remote control, the control obtains the user identity from the token and instructs the TV set to list the user's favorite channels. The user can also touch other smart objects to personalize them, e.g., touch a music player for the favorite music, switch on a light that automatically tunes to the user's favorite color temperature or hue, etc.

If the user can protect the personal wearable token well, the touch-to-access device authentication can also be used in more access-critical scenarios. For example, a touch on a smartphone or tablet unlocks the device's screen automatically, allows in-app purchases, passes the parental controls, etc. Beyond the above use scenarios for improved convenience in access control, the touch-to-access scheme can also enhance the security of various systems. For instance, it can be used with fingerprint scanning to form a two-factor authentication against fake fingerprints. A wireless reader can access a worn medical sensor only if the reader has a physical contact with the wearer's skin. The contact enforces the wearer's awareness regarding the access and prevents remote wireless attacks with stolen credentials \cite{halperin2008pacemakers}. {\rev Thus, the touch-to-access scheme will be more secure than the existing hardware token approaches such as Duo \cite{duo}.}

The essence of the touch-to-access scheme is the detection of whether the wearable token and the smart object in question have physical contact with the same user's body. Existing studies tackle this same-body contact detection problem by intra-body communication (IBC) \cite{zimmerman1995personal,matsushita2000wearable,park2006tap,baldus2009human,vu2012distinguishing,holz2015biometric,touchpairing} and physiological sensing such as
electrocardiography (ECG) \cite{poon2006novel,imdguard,pska,opfka,h2h}, photoplethysmogram (PPG) \cite{poon2006novel,pska,opfka}, and electromyogram (EMG) \cite{SecretfromMuscle}. IBC requires either non-trivial customized transceivers \cite{zimmerman1995personal,matsushita2000wearable,park2006tap,baldus2009human,touchpairing} or a touchscreen as the receiver \cite{vu2012distinguishing,holz2015biometric}, resulting in increased cost or reduced applicable scope. The physiological sensing approaches are based on a {\em body-area property}, i.e., the physiological signals captured from the same human body have similar values or features, whereas those collected from different human bodies are distinct. However, the physiological sensors are often bulky due to the required physical distances among a sensor's electrodes \cite{barill2003ecg,SecretfromMuscle}. Furthermore, they often need careful placement and may perform poorly in daily life settings \cite{chang2012body}.

Different from IBC and physiological sensing, in this paper, we investigate the feasibility and effectiveness of using {\em induced body electric potential} (iBEP) due to the {\em body antenna effect} for touch-to-access device authentication. As a non-physiological phenomenon, the body antenna effect refers to the alteration of the intensity of the mains hum captured by an analog-to-digital converter (ADC) when the ADC has a physical contact with a human body. The mains hum induced by the building's electrical cabling is ubiquitous. In addition, ADC is a basic electronic component that is widely available on microcontrollers. Recent studies have exploited the body antenna effect for key stroke detection \cite{elfekey2013design}, touch sensing \cite{cohn2011your}, motion detection \cite{cohn2012ultra}, gesture recognition \cite{humantenna}, and wearables clock synchronization \cite{yan2017application}. These studies leverage several characteristics of iBEP, such as signal intensity  alteration \cite{elfekey2013design} and periodicity \cite{yan2017application}, or feed iBEP signals to machine learning algorithms for motion and gesture recognition \cite{cohn2011your,cohn2012ultra,humantenna}.
Differently, to use iBEP for device authentication, {\rev its body-area property and the underlying physical mechanism need to be well understood.}
To the best of our knowledge, these issues have not been studied.

In this paper, we discuss in detail the physical mechanism of the iBEP's generation and its body-area property.
The iBEP measurement by an ADC is the difference between the electric potentials of the ADC pin and the ground\footnote{Throughout this paper, ``ground'' refers to the floating ground of a device.} of the sensor, respectively. From electrostatics, a human body, which can be viewed as an uncharged conductor, will alter its nearby electric field (EF) emitted from the electrical cabling of the building due to electrostatic induction.
As a result, the iBEP measurement by a sensor will be affected by the presence of a nearby human body. In particular, we make the following two hypotheses
based on the above understanding.
First, the iBEP signals measured by two sensors that are on the same human body and close to each other
will be similar.
This is because 1) the two sensors' ADC pins will have the same potential due to their connections to the equipotential human body,
and 2) their grounds will most likely have similar potentials as they are close to each other in the EF.
Second, the iBEP signals collected from different human bodies will be different. This is because different human bodies will most likely have different potentials and thus affect nearby EFs differently since they build up different surface charge distributions in the electrostatic induction.

Our extensive measurement results are consistent with the above two hypotheses.
  Based on the results, we design a prototype system called {\em TouchAuth} that performs touch-to-access device authentication based on iBEP signals.
We implement the same-body contact detection algorithm based on two similarity metrics, i.e., absolute Pearson correlation coefficient (APCC) and root mean square error (RMSE).
Extensive experiments show that the APCC-based TouchAuth achieves true acceptance rates of 94.2\% and 98.9\% subject to a false acceptance rate upper bound of 2\% when one and five seconds of iBEP signal is recorded, respectively. In contrast, ECG/PPG approaches \cite{poon2006novel,pska,opfka} need to record the signal(s) for tens of seconds to achieve comparable detection accuracy (cf.~\sect\ref{sec:related}).
Our experiments also show that various possible interfering sources including appliances' electromagnetic emanations and noise injections into power networks do not affect TouchAuth.

In summary, TouchAuth is a low-cost, lightweight, and convenient approach for the authorized users to access smart objects in indoor environments. To implement TouchAuth, the smart object's and the wearable token's microcontroller ADCs are to be wired to their conductive exteriors. Compared with the near-field communication (NFC) approach that enforces a proximity requirement on device authentication, the touch requirement of TouchAuth is more intuitive and clearer. Moreover, compared with the ADCs that are widely available on microcontrollers, the NFC chips are more costly and need to be integrated into the smart objects to read the wearable tags.

{\rev
The contributions of this paper are summarized as follows:
\begin{itemize}
  \item We explain the generation mechanism of iBEP and show that iBEP is an effective signal for devising a touch-to-access device authentication approach.
  \item We design an iBEP-based device authentication approach called TouchAuth. It uses iBEPs to detect whether two devices are in proximity on the same human body.
\item Extensive experiments under real-world settings are conducted to evaluate the performance of TouchAuth.
\end{itemize}
}

The rest of this paper is organized as follows. 
{\rev \sect\ref{sec:sys-ove} presents the system and threat models, and the approach overview.} \sect\ref{sec:basis} presents the electrostatics of iBEP generation and states the objective of this paper. \sect\ref{sec:measurement} presents the measurement study. \sect\ref{sec:approach} and \sect\ref{sec:eval} design and evaluate TouchAuth, respectively. \sect\ref{sec:limit} discusses several issues. \sect\ref{sec:related} reviews related work. \sect\ref{sec:conclude} concludes the paper.

\section{System Overview}
\label{sec:sys-ove}

\subsection{System Model}
\label{subsec:sys-model}

We consider an authentication system with two devices that have been previously paired, i.e., an {\em authenticator} and an {\em authenticatee}. We assume that the two devices have a wireless communication channel, e.g., Wi-Fi, Bluetooth (Low Energy), Zigbee, etc. The pairing enables them to communicate. The authenticator is a trustworthy device that can sense the iBEP signal $s(t)$, $\forall t$, at a location $\mathcal{L}$ on the body of a user $\mathcal{U}$. To be authenticated, the authenticatee presents its sensed iBEP signal $s'(t)$, $t \in [t_1, t_2]$, to the authenticator. The $\ell = t_2 - t_1$ is called {\em signal length}. The authenticatee is {\em valid} only if it has physical contact with a location $\mathcal{L}'$ on $\mathcal{U}$ which is close to $\mathcal{L}$ such that $s'(t) \approx s(t)$, $\forall t \in [t_1, t_2]$; otherwise, it is {\em invalid}. The valid authenticatee will be granted a certain access; the invalid authenticatee will be denied the access. We assume that the clocks of the authenticator and the authenticatee are synchronized, such that the authenticator can select a segment of $s(t)$ in the time duration $[t_1, t_2]$ to check the similarity between $s(t)$ and $s'(t)$ for same-body contact detection.
Before the authentication process, clock synchronization can be achieved using existing approaches \cite{maroti2004flooding,li2012flight,yan2017application}.

We now discuss the roles of different devices in the scenarios discussed in \sect\ref{sec:intro}. When the user with a wrist wearable token touches a smartphone to unlock its screen, the wearable token is the authenticator, whereas the automatic unlock program on the phone is the authenticatee.
On detecting human touch (by either button/touchscreen press or increased iBEP intensity), the unlock program presents its captured iBEP signal to the wearable token that will perform the same-body detection. A positive detection result allows the program to unlock the smartphone; otherwise, the program should not unlock the phone.
In the example of worn medical sensor access, the medical sensor is the authenticator, whereas the wireless reader is the authenticatee. Only the reader that has physical contact with the sensor wearer will receive a one-time password to access the data on the sensor.
In the less access-critical examples of personalizing smart objects, the wearable token (i.e., the authenticator) transmits the user's identity to the touched smart object (i.e., the valid authenticatee) for personalization.

\subsection{Threat Model}
\label{subsec:threat-model}

We adopt the same threat model that is used for an ECG-based device authentication system in \cite{h2h}. Specifically, we consider an adversary who fully controls the communication channel between the authenticator and any valid authenticatee and aims at impersonating the valid authenticatee. The channel control includes eavesdropping, dropping, modifying, and forging messages as desired. The adversary can corrupt neither the authenticator nor the valid authenticatee.

\subsection{Approach Overview}
\label{subsec:approach-overview}

\begin{figure}
  \centering
\includegraphics{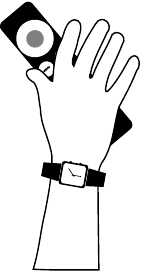}
  \hspace{2em}
    \includegraphics{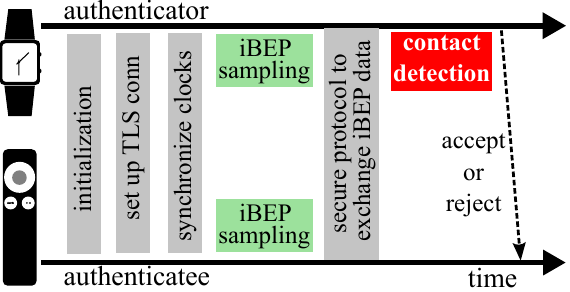}
    \caption{Left: A use scenario where the smart watch personalizes a remote control and the associated media system by a touch; Right: authentication process.}
    \label{fig:overview0}
\end{figure}

Fig.~\ref{fig:overview0} illustrates an authentication process of our approach. The authentication process can be initiated by the authenticatee upon it detects a human touch based on iBEP. After a handshake with a nearby authenticator, a Transport Layer Security (TLS) connection is set up between the authenticator and the authenticatee to ensure data confidentiality, integrity, and freshness of consequent communications. TLS is feasible on mote-class platforms \cite{fouladgar2006tiny,schmitt2017two}. Because the authenticatee's certificate presented during the TLS setup needs not to be validated by the authenticator, our approach does not involve a cumbersome public key infrastructure (PKI). Then, the two parties synchronize their clocks and sample their respective iBEPs $s(t)$ and $s'(t)$ synchronously for $\ell$ seconds. After that, following an existing protocol H2H \cite{h2h} that is designed for ECG-based device authentication, the two parties perform a commitment-based data exchange to ensure the security of the system against the threat defined in \sect\ref{subsec:threat-model}. Note that, without using H2H, a naive approach of transmitting $s'(t)$ from the authenticatee to the authenticator over the TLS connection for contact detection is vulnerable to a man-in-the-middle attack based on full channel control \cite{h2h}. After obtaining $s'(t)$, the authenticator runs a same-body contact detection algorithm with $s(t)$ and $s'(t)$ as inputs to decide whether the authenticatee is valid. Lastly, the authenticator notifies the authenticatee of acceptance or rejection.

Note that in the less security-critical use scenarios such as smart object personalization, the TLS connection setup can be skipped and the commitment-based data exchange procedure can be replaced with a normal data exchange procedure. This reduces overhead.

\section{Basis and Research Objective}
\label{sec:basis}

{\rev In this section, we discuss the physical basis of TouchAuth (\sect\ref{subsec:body-antenna}) and the research objective (\sect\ref{subsec:objective}).}

\subsection{Body Antenna Effect}
\label{subsec:body-antenna}
      
\begin{figure}[t]
  \centering
  \includegraphics[width=\columnwidth]{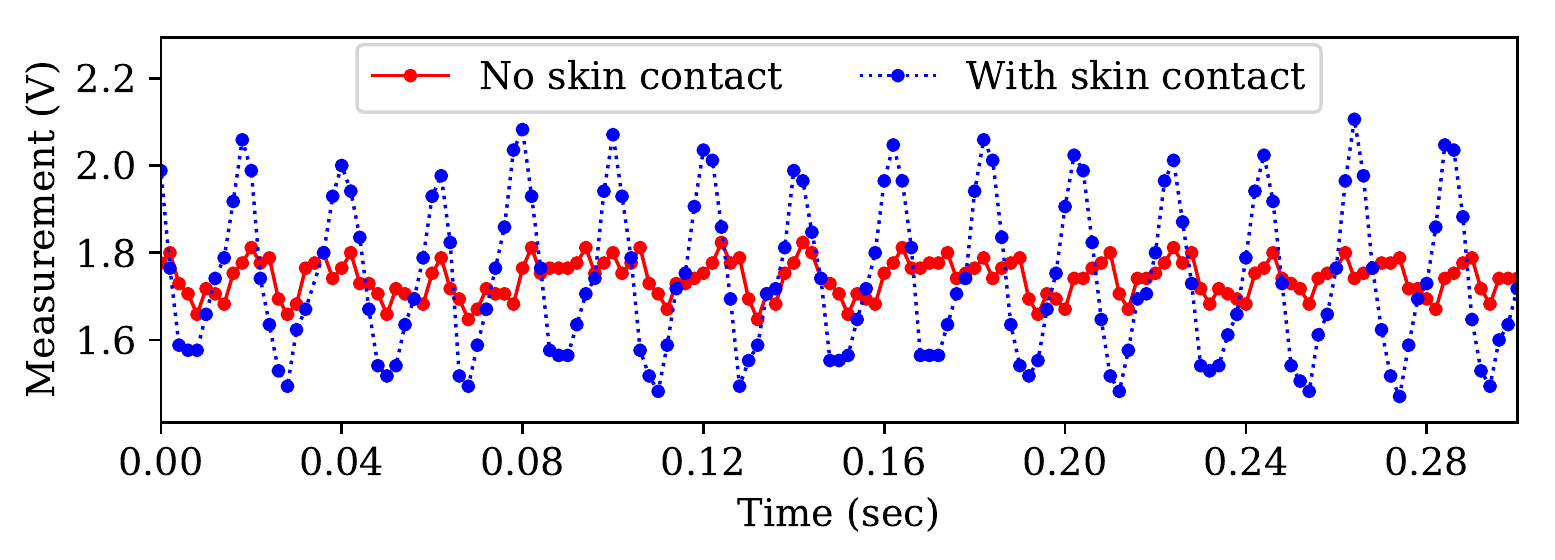}
  \caption{The body antenna effect.}
  \label{fig:bep}
\end{figure}

First, we illustrate the body antenna effect. The two curves in Fig.~\ref{fig:bep} are the measurement traces of a mote-class sensor placed at a fixed position, with an ADC pin floating in the air or pinched by a person, respectively. More details of the sensor will be presented in \sect\ref{subsec:measurement-setup}. Without body contact, the sensor captures the mains hum with weak amplitude and a frequency of about $50\,\text{Hz}$ (i.e., the nominal grid frequency in our region). With body contact, the signal has greater amplitude and exhibits more clearly the frequency of $50\,\text{Hz}$. The above result shows that the human body affects the reception of mains hum. Several recent studies \cite{humantenna,elfekey2013design,yan2017application} exploited this human body antenna effect for various applications. However, they do not provide an in-depth explanation of the effect. This section explains this effect in detail, which will guide our experiments and the design of TouchAuth.

\subsubsection{Electric field (EF) from electrical cabling}

A line of charge emits an EF, whereas a current through the line generates a magnetic field. Thus, a charged wire carrying alternating current (ac) will generate both time-varying electric and magnetic fields. However, since the magnetic fields generated from the two close-lying current-carrying wires within a single power cable tend to cancel each other, the overall magnetic field around a power cable is normally very weak \cite{radiation-home}.\footnote{The low-intensity net magnetic field due to a small mismatch between the two wires' currents, which is caused by the vagabond currents effect, is sensible using special devices such as hall effect sensors and tank circuits tuned to the power grid frequency.}
At the nominal frequency of the ac power grid, i.e., $50$ or $60\,\text{Hz}$, the power cable's EF is an extremely low frequency (ELF) radiation with a wavelength of thousands of miles. At such a wavelength scale, we do not need to consider the magnetic field excited by the time-varying EF. Thus, EF is the main emanation from a power cable.

Modern buildings often have complex electrical cabling. Permanent power cables run above ceilings, below floors, on walls, etc. There are also power extension cords installed by residents. As the EF from a cable is a vector field with intensity attenuating with the distance from the cable, the combined EF caused by all the cables in a building is a vector field with a complex intensity distribution over the space. In normal homes with $220\,\text{V}$ power supply, the intensity of the combined EF is often between $3\,\text{V/m}$ and $30\,\text{V/m}$ \cite{radiation-home}.
{\rev
The EF is the superposition of the EFs emitted by surrounding electrified power cables and appliances. Due to the spatial distribution of the power cables and appliances, the gradients of the EF at different locations are generally different.
}

\begin{figure}
  \centering
  \vspace{-1em}
  \subfigure[Heater powered on and off.]
  {
    \includegraphics[width=.47\columnwidth]{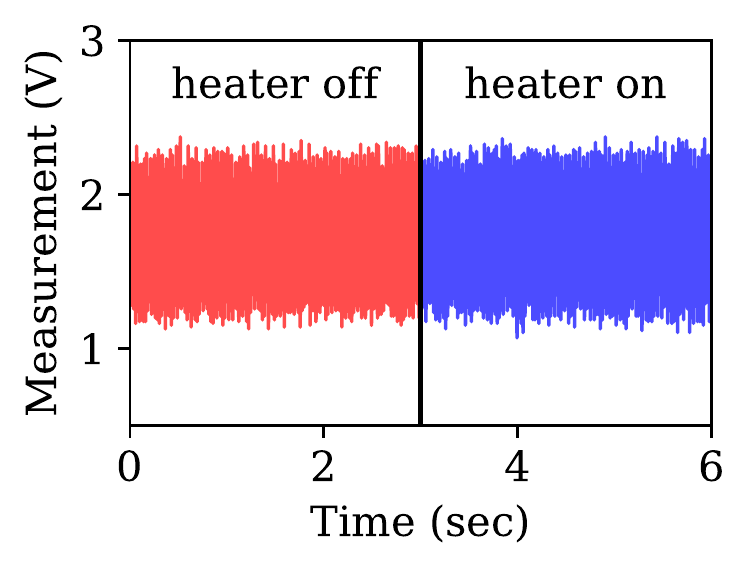}
    \label{fig:heater}
  }
  \hfill
  \subfigure[Cord (dis)electrified.]
  {
    \includegraphics[width=.47\columnwidth]{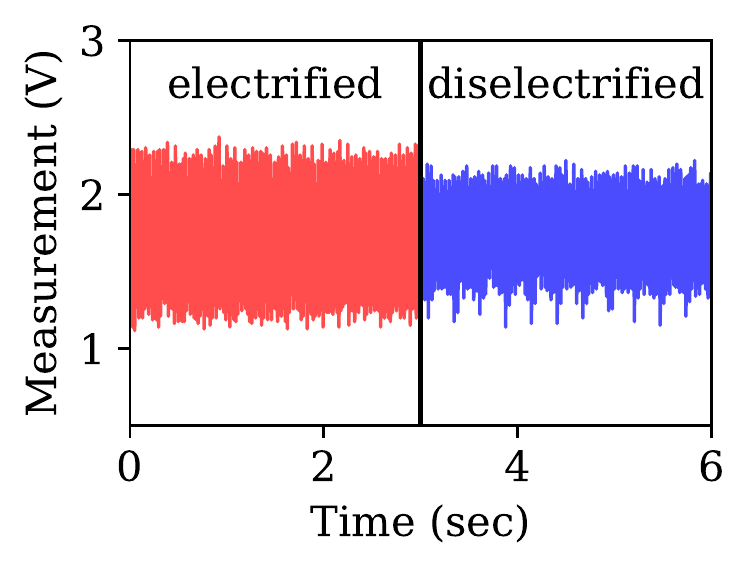}
    \label{fig:cord}
  }
  \caption{Mains hum measured by a sensor (without human body contact) placed close to a power cord supporting a $2\,\text{kW}$ heater.}
  \label{fig:mains-hum-app}
  \vspace{-0.5em}
\end{figure}

Note that if an appliance is powered off, the intensity of the EF from the power cable supporting the appliance will remain unchanged. This is due to the fact that most switches only break the connection in one wire, while the wires will still have the same service voltage as when the appliance is powered on. We conduct an experiment to verify this. Fig.~\ref{fig:heater} shows the mains hum measured by a sensor that has a conductor wire connected to an ADC pin to improve EF sensing and is placed close to a power cord supporting a $2\,\text{kW}$ heater. From the figure, the operating status of the heater does not affect the measurements. Fig.~\ref{fig:cord} shows the sensor's measurements when the heater remains off and the power cord is connected to or disconnected from the wall outlet. We can see that the intensity of the sensor readings is weaker when the power cord is diselectrified. The remaining intensity is caused by the EFs from other electrified power cables in the building. The above results suggest that (i) the ambient field is an EF caused by the ac voltages, (ii) the ac current changes caused by appliances' operating status changes have little impact on the ambient field.

\subsubsection{Interactions among EF, sensor, and human body}
\label{subsubsec:interaction}

First, we discuss the situation without a human body. Our discussions below concern a time instant only.
As a typical sensor's ADC has high input impedance (hundreds of $\text{k}\Omega$ up to a few $\text{M}\Omega$), the ADC pin and the ground of the sensor can be considered insulated for simplicity of discussion. The ADC and ground will have different potentials in the building's ambient EF due to their physical distance. The potential difference is the measurement of the sensor. For instance, in an EF with an intensity up to $30\,\text{V/m}$, if the equivalent distance between the ADC pin and the ground is $1\,\text{cm}$, the measurement can be up to $0.3\,\text{V}$. This is consistent with our results in Figs.~\ref{fig:bep} and \ref{fig:mains-hum-app}.

Now, we discuss the situation with a human body. A body can be viewed as a conductor due to its low impedance (a few $\text{k}\Omega$ \cite{reilly1998applied}). From electrostatics, an uncharged conductor in an EF will build up a surface charge distribution to reach an electrostatic equilibrium, where the EF inside the conductor is zero and the conductor's surface is an equipotential surface \cite{purcell2013electricity}. The surface charge distribution will generate an EF. As a result, the EF combining that from the original source (i.e., electrical cabling) and the electrostatically induced conductor (i.e., the human body) is different from the EF in the absence of the conductor. In other words, the human body affects its nearby EF. The change of field intensity results in the change of potential difference between the sensor's ADC and ground, i.e., the sensor's measurement.

\begin{figure}
  \centering
  \includegraphics[width=.6\columnwidth]{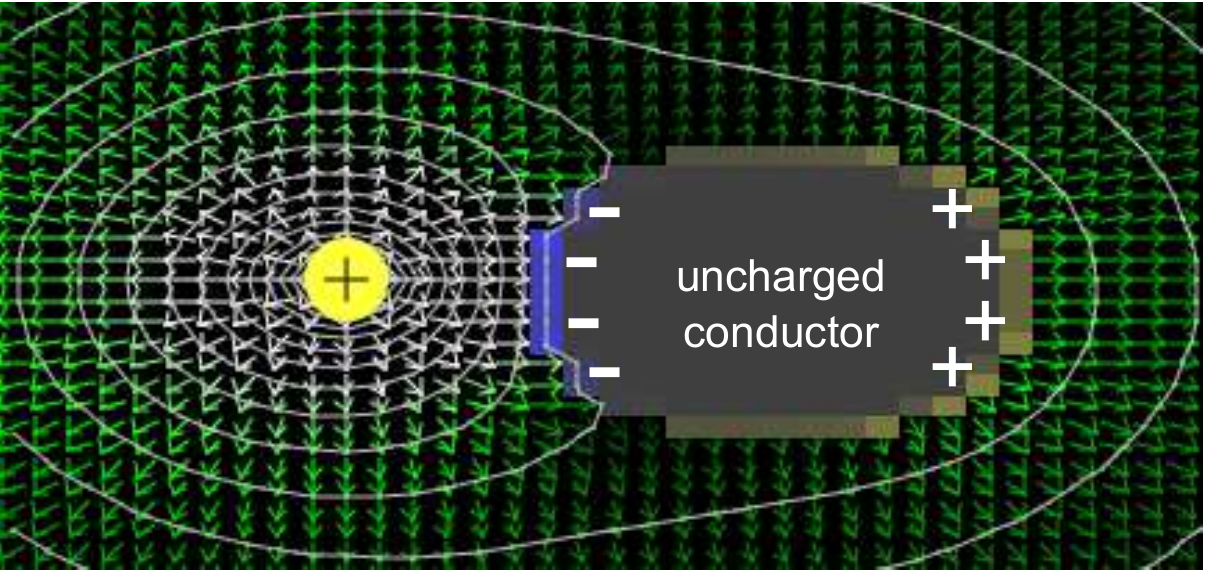} 
  \caption{Impact of an uncharged conductor (e.g., human body) on an EF from a point charge. White curves represent equipotential lines. Image credit: Paul Falstad \cite{electrostatics-tool}.}
  \label{fig:body-field}
\end{figure}

We use a 2D electrostatics simulator \cite{electrostatics-tool} to generate an example as shown in Fig.~\ref{fig:body-field} that illustrates the human body's impact on ambient EF.
When an uncharged conductor is in the field, negative/positive surface charges will be built up. As a result, the EF intensity, which is characterized by the density of the equipotential lines, will change in the space close to the electrostatically induced conductor. For instance, in Fig.~\ref{fig:body-field}, the EF between the charge and the conductor is intensified. The reading of a sensor in this area will increase if its ADC and ground are arranged in the direction of the field. In practice, the indoor EF will be much more complex than the one shown in Fig.~\ref{fig:body-field}. Nevertheless, the example provides a basic understanding of the body antenna effect.

\subsection{Research Objective}
\label{subsec:objective}

As discussed in \sect\ref{subsubsec:interaction}, the human body in an EF is an equipotential conductor. Considering two sensors with their ADC pins connected to the same human body, the potentials of their ADC pins will be the same. If they are close to each other, their grounds will have similar potentials. Thus, their readings will be similar. If the two sensors are attached to two locations on the human body which are far from each other, their grounds will have different potentials. As a result, though their ADC pins have the same potential due to the human body contact, their readings will be different.

Now, we discuss the case where the two sensors are on different human bodies. The human bodies will most likely have different potentials. Moreover, even if we ignore the impact of the two human bodies on the EF, because the two sensors are at two different locations, the gradients of the indoor EF at the two locations will be most likely different. As a result, the two sensors' measurements will be different. This difference will be further intensified by the different impacts of the two human bodies on their nearby EFs.

Our research objective is two-fold. First,
we aim to verify the above inferences from the iBEP electrostatics via an extensive measurement study, which is the subject of \sect\ref{sec:measurement}. If the measurement results are supportive of the inferences, we will inquire whether iBEP sensing can be exploited to implement the desirable touch-to-access scheme. This will be addressed in \sect\ref{sec:approach} and \sect\ref{sec:eval}.

\section{Measurement Study}
\label{sec:measurement}

\subsection{Measurement Setup}
\label{subsec:measurement-setup}

\begin{figure}
  \centering
  \begin{minipage}[t]{.25\columnwidth}
	\includegraphics[width=\textwidth]{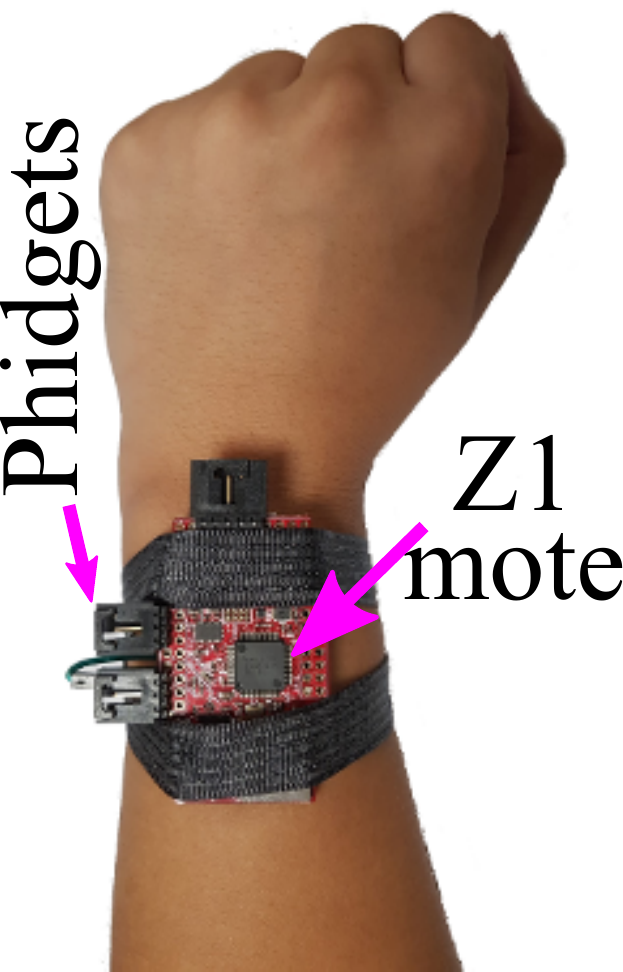}
    \caption{Z1.}
    \label{fig:z1}
  \end{minipage}
  \hfill
  \begin{minipage}[t]{.65\columnwidth}
	\includegraphics[width=\textwidth]{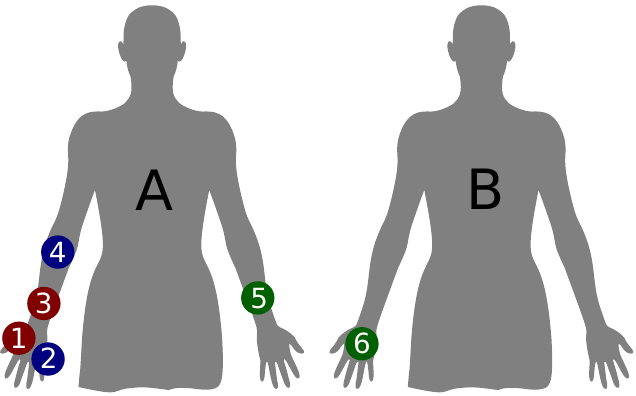}
    \caption{Sensor placements.}
    \label{fig:placements}
  \end{minipage}
  \end{figure}

Our experiments are conducted using several Zolertia Z1 motes \cite{z1} and a Kmote \cite{kmote}. Both types of motes are equipped with MSP430 microcontroller and CC2420 802.15.4 radio. The Z1 motes are used to collect iBEP data from human bodies, whereas the Kmote is used as a base station to synchronize the Z1 motes' clocks and collect their iBEP data over wireless. Each Z1 mote is powered by a lithiumion polymer battery; the Kmote base station is connected to a desktop computer through a USB cable. Each Z1 mote has two Phidgets sensor ports connected to several ADC pins of its microcontroller. We use a conductive wire as an electrode to create a physical contact between a pin in one Phidgets sensor port and the skin of the Z1 wearer. Fig.~\ref{fig:z1} shows a Z1 worn on a wrist. The motes run TinyOS 2.1.2. The program running on the Z1 mote samples the ADC at a rate of $500\,\text{sps}$. The samples are timestamped using the Z1's clock. The program uses a reliable transmission protocol called Packet Link Payer \cite{pll} to stream the samples to a Kmote base station. It also integrates the Flooding Time Synchronization Protocol (FTSP) \cite{maroti2004flooding} to synchronize the Z1's clock to the Kmote base station.

\subsection{Measurement Results}
\label{subsec:measurement-results}

We conduct three sets of experiments in a lab office.

\subsubsection{Insensitivity to time-varying magnetic field (MF)}
\label{subsubsec:insensitivity-mf}

\begin{figure}
  \hfill
  \includegraphics[width=0.38\columnwidth]{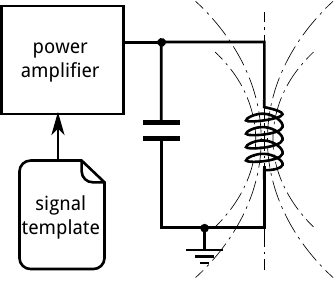}
  \hfill
  \includegraphics[width=0.4\columnwidth]{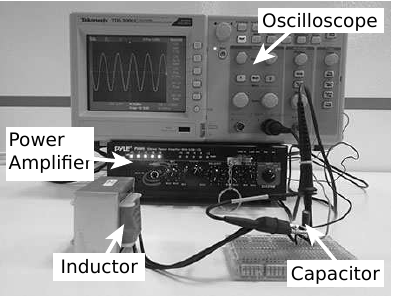}
  \hfill
  \caption{Time-varying magnetic field generator.}
  \label{fig:injector}
\end{figure}

To verify that the body antenna effect is mainly caused by EF, rather than MF, we build an MF generator and examine its impact on iBEP sensing. Fig.~\ref{fig:injector} shows the schematic and the implemented MF generator. It consists of a {\em power amplifier} that weighs $2.6\,\text{kg}$, a $320\,\text{mH}$ inductor, and a $1\,\mu\text{F}$ capacitor. The power amplifier admits a specified signal waveform and outputs the corresponding current to induce the inductor to generate time-varying MF. The capacitor is used to smooth the output signal during the induction. In this experiment, the specified signal is a $85\,\text{Hz}$ sinusoid. {\rev 
The choice of this frequency has two reasons: a) As $85\,\text{Hz}$ is close to the grid frequency of $50\,\text{Hz}$ in our region, the iBEP sensor will have similar signal reception performance as for the $50\,\text{Hz}$ signal from powerlines; b) The choice of $85\,\text{Hz}$ is also to avoid the harmonics of the grid frequency, i.e., $100\,\text{Hz}$, $150\,\text{Hz}$ and so on.} We configure the power amplifier to use its maximum gain. From our tests, this generator causes strong interference to nearby tank circuits that can sense MF changes. However, from our experiments, it generates little impact on nearby on-body Z1-based iBEP sensors. From a frequency analysis, the power density of the iBEP signal at $85\,\text{Hz}$ is 66 times weaker than that at $50\,\text{Hz}$ when the on-body iBEP sensor is only $2\,\text{cm}$ away from the inductor. {\rev The intensity of the signal at $85\,\text{Hz}$ is similar to that of the ambient noise. The reason is that the iBEP sensor (i.e., an ADC with floating ground) is an open circuit, which cannot be induced by the time-varying magnetic field.}
This result confirms that the body antenna effect is mainly caused by EF.

\subsubsection{iBEPs on the same body}
\label{subsubsec:h1-test}

\begin{figure}
  \centering
  \subfigure[Measurements of two sensors in two seconds.]
  {
    \includegraphics[width=\columnwidth]{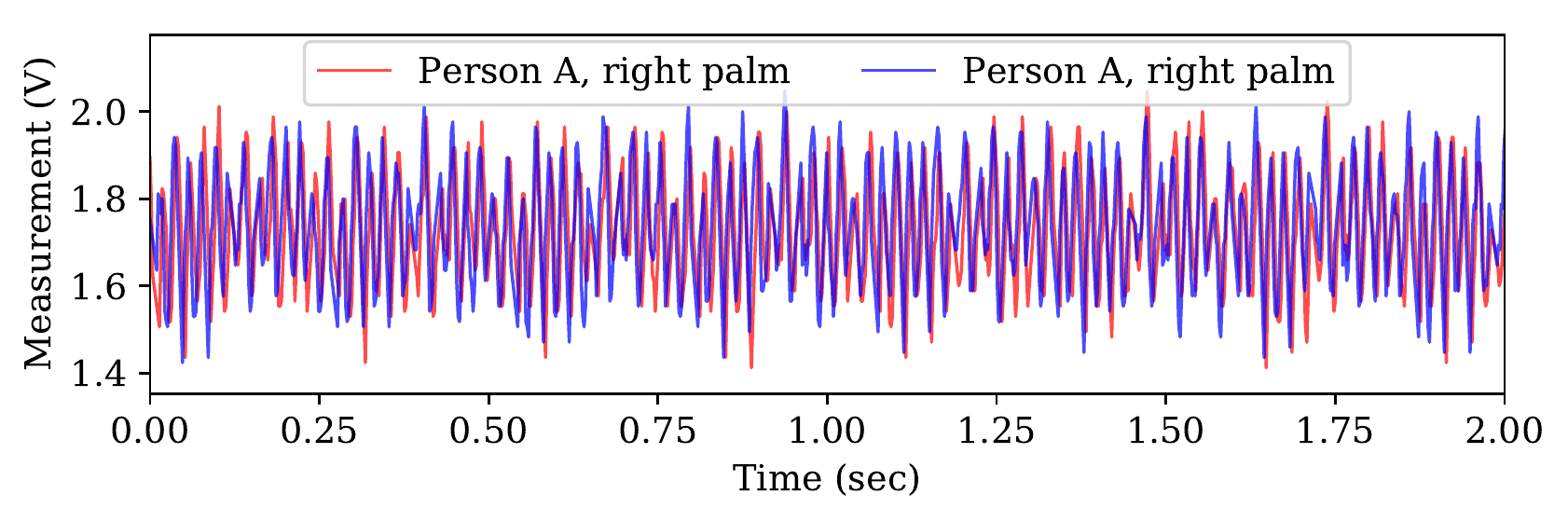}
    \label{fig:same-body-same-loc}
  }
  \hfill
  \subfigure[Zoomed-in view.]
  {
    \includegraphics[width=\columnwidth]{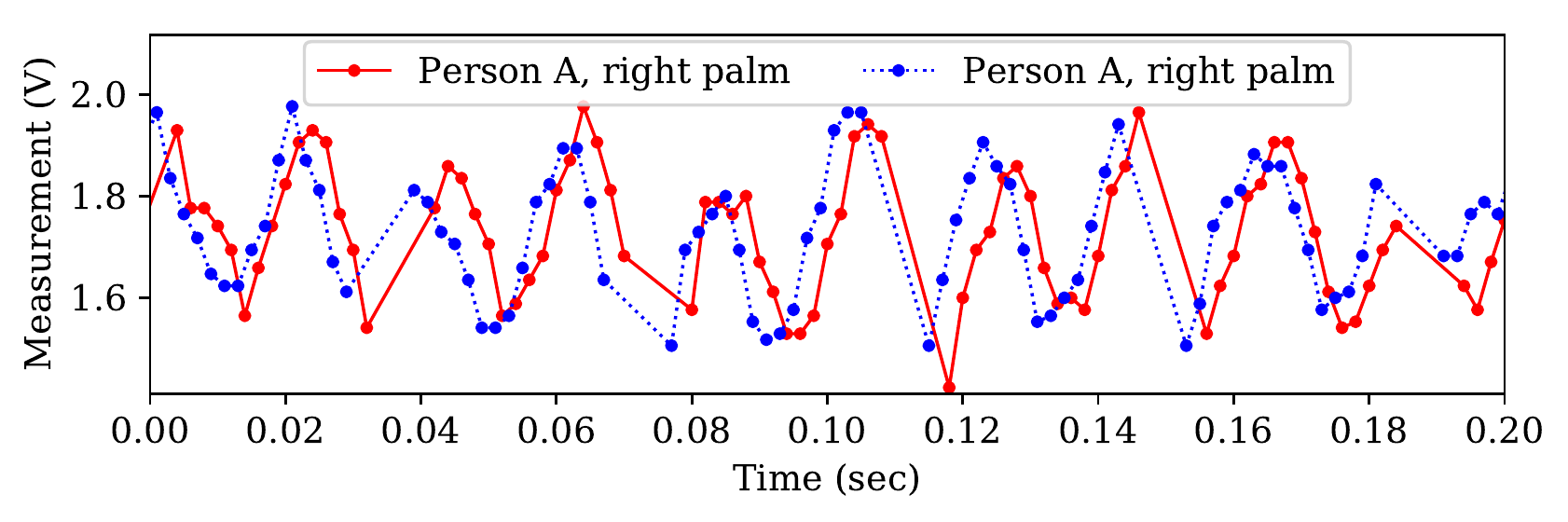}
    \label{fig:same-body-same-loc-zoom-in}
  }
  \caption{iBEPs measured by two sensors in the same palm when the holder sits in a chair.}
  \label{fig:same-hand}
  \end{figure}

First, Person~A sits in a chair and uses his right hand palm to hold two Z1 sensors steadily. The ADCs of both sensors have direct contact with the palm skin. In Fig.~\ref{fig:placements}, the nodes numbered \ding{182} and \ding{183} illustrate the placement of the two sensors. Fig.~\ref{fig:same-body-same-loc} shows the iBEPs captured by the two sensors over two seconds. Fig.~\ref{fig:same-body-same-loc-zoom-in} shows a zoomed-in view of Fig.~\ref{fig:same-body-same-loc}. From the two figures, we can see that the two iBEP signals are synchronous and of the same amplitude level. This shows that, when the two sensors are in proximity on the same human body, their measurements are similar.

\begin{figure}
  \centering
  \subfigure[Spot X.]
  {
    \includegraphics[width=.47\columnwidth]{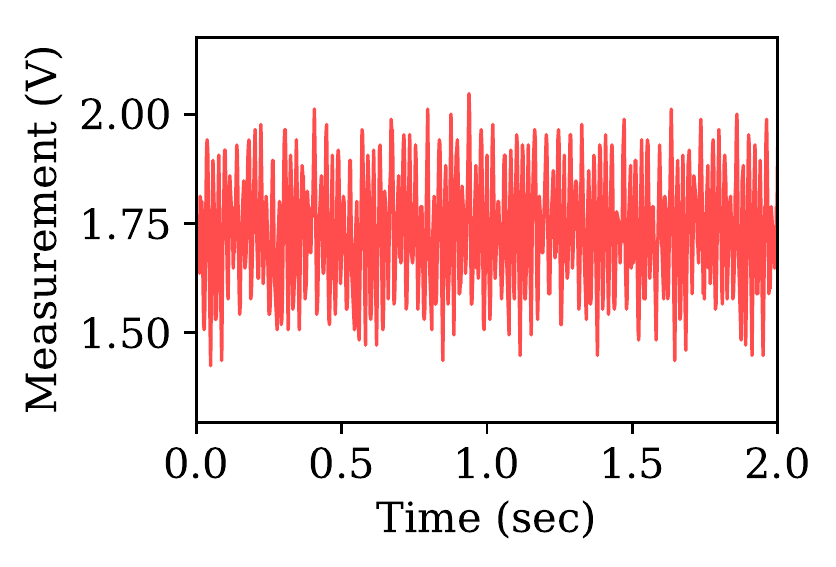}
    \label{fig:same-body-diff-loc-a}
  }
  \hfill
  \subfigure[Spot Y.]
  {
    \includegraphics[width=.47\columnwidth]{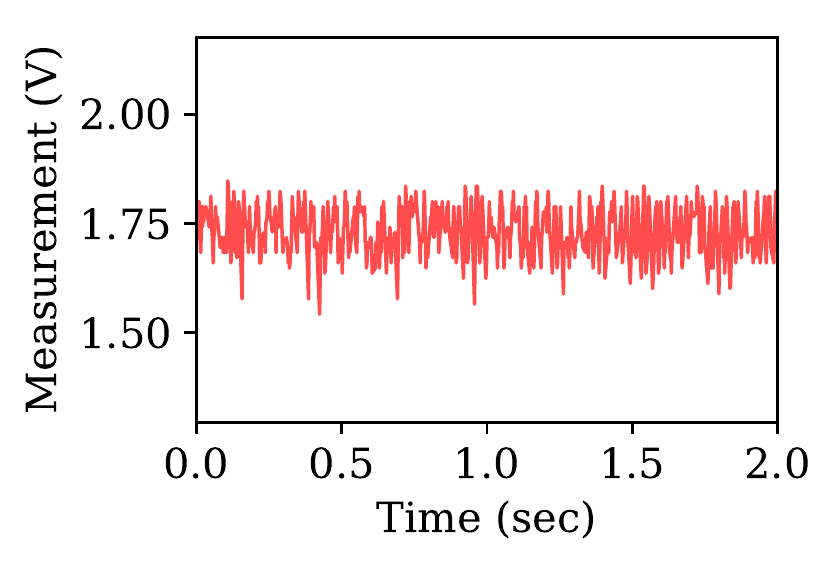}
    \label{fig:same-body-diff-loc-b}
  }
  \caption{iBEPs measured by a sensor in the same palm when the wearer stands at different spots in the lab.}
  \end{figure}

Second, we investigate the impact of spatial location on iBEP. As discussed in \sect\ref{subsec:body-antenna}, the indoor EF has an intensity distribution over space. Thus, the potential difference between the human body and the ground of the sensor will vary with location.
In this experiment, Person~A holds a sensor in his palm with skin contact and stands at two spots in the lab. Fig.~\ref{fig:same-body-diff-loc-a} and Fig.~\ref{fig:same-body-diff-loc-b} show the iBEPs at the two spots that are about one meter apart. From the two figures, the amplitude of the iBEP at Spot~X is larger than that at Spot~Y. Note that Spot~X is closer to a cubicle with a number of electrified power cables and power extensions.

\begin{figure}
  \centering
  \subfigure[Two nodes on the right arm.]
  {
    \includegraphics[width=.47\columnwidth]{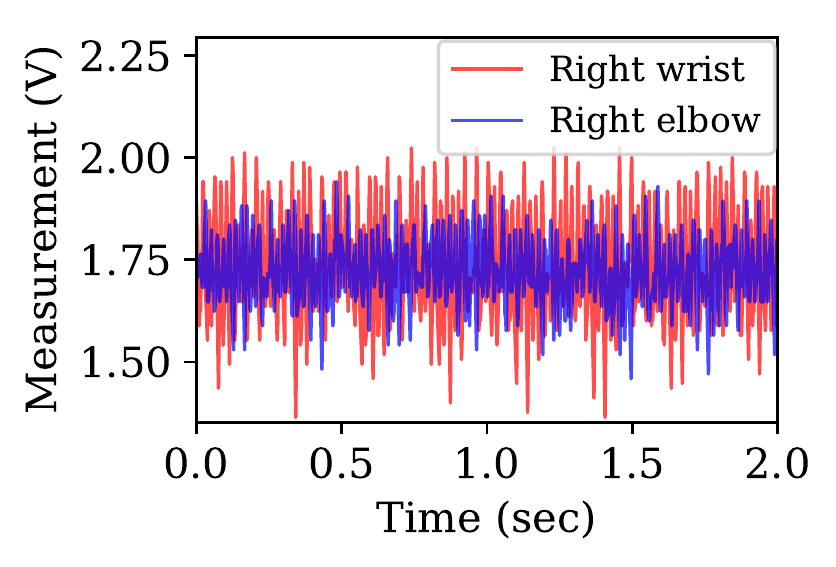}
    \label{fig:same-body-same-arm}
  }
  \hfill
  \subfigure[Two nodes on different arms.]
  {
    \includegraphics[width=.47\columnwidth]{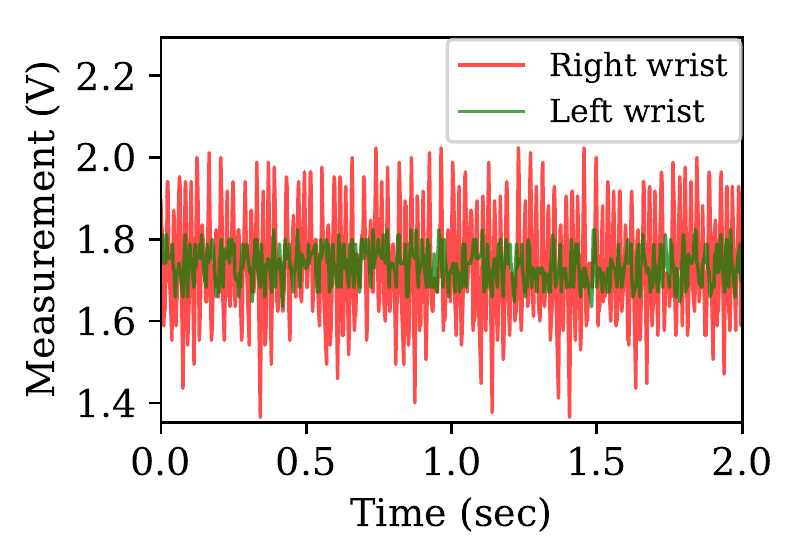}
    \label{fig:same-body-diff-arm}
  }
  
  \subfigure[Zoomed-in view.]
  {
    \includegraphics[width=\columnwidth]{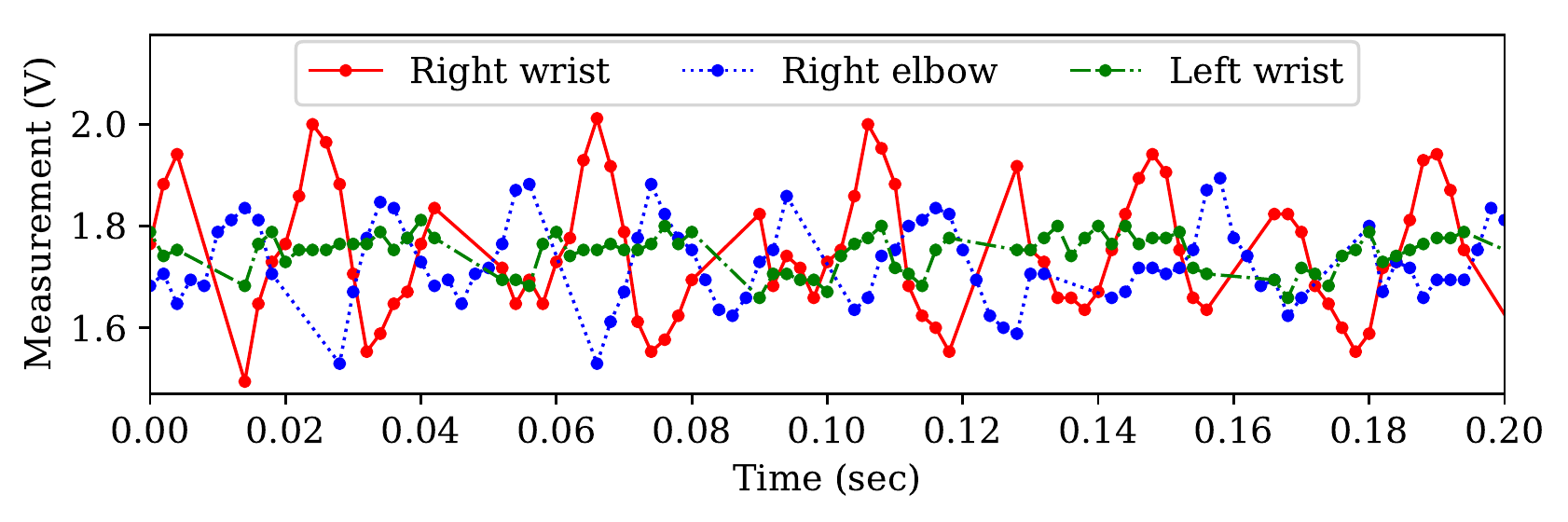}
    \label{fig:same-body-arms-zoomed-in}
  }
  \caption{iBEPs at different locations of Person A.}
  \label{fig:same-body-diff-arms-all}
  \end{figure}

Third, we investigate the impact of the sensor placement on the received iBEP signal. We place three sensors on Person~A, two on the right arm and the remaining one on the left arm. The two sensors on the right arm are separated by about $15\,\text{cm}$, one of which is close to the wrist and the other is close to the elbow. In Fig.~\ref{fig:placements}, the nodes numbered \ding{184}, \ding{185}, and \ding{186} illustrate the placement of the three sensors. In this experiment, the person stands and keeps a side lateral raise posture.
Fig.~\ref{fig:same-body-diff-arms-all} shows the iBEP signals collected from the three sensors in the same time period. Fig.~\ref{fig:same-body-same-arm} shows the iBEPs measured by the two sensors on the right arm. Fig.~\ref{fig:same-body-diff-arm} shows the iBEPs measured by two sensors on different arms. Fig.~\ref{fig:same-body-arms-zoomed-in} shows the zoomed-in view for the signals in Figs.~\ref{fig:same-body-same-arm} and \ref{fig:same-body-diff-arm}. From the results, we can see that the signals measured by the two sensors on the right arm have similar amplitudes, but a phase shift of about $180^{\circ}$. This can be caused by that the ADC-to-ground directions of the two sensors in the EF are different. Ignoring the phase shift, the signals measured by the two sensors $15\,\text{cm}$ apart on the same arm exhibit higher similarity than those measured by the two sensors on different arms, but lower similarity than those measured by the two sensors in the same palm as shown in Fig.~\ref{fig:same-body-same-loc-zoom-in}.

\begin{figure}
  
  \centering
  \begin{minipage}[t]{0.49\columnwidth}
    \centering
	\includegraphics[width=\textwidth]{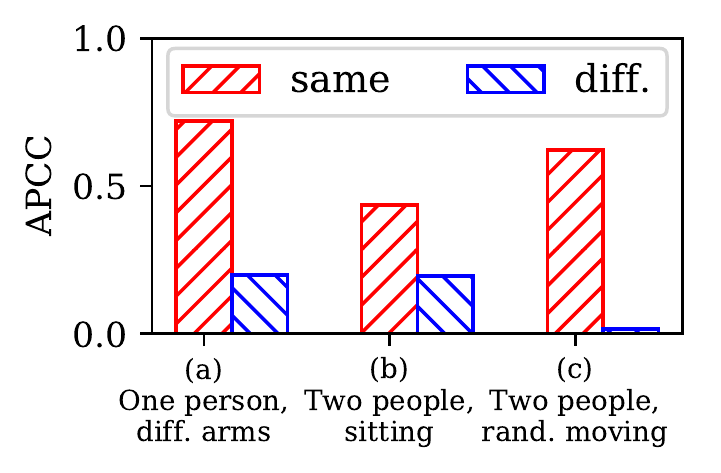}
    \caption{\rev APCC in different placements.}
    \label{fig:pearson}
  \end{minipage}
  \hfill 
  \begin{minipage}[t]{0.49\columnwidth}
    \centering
	  \includegraphics[width=\textwidth]{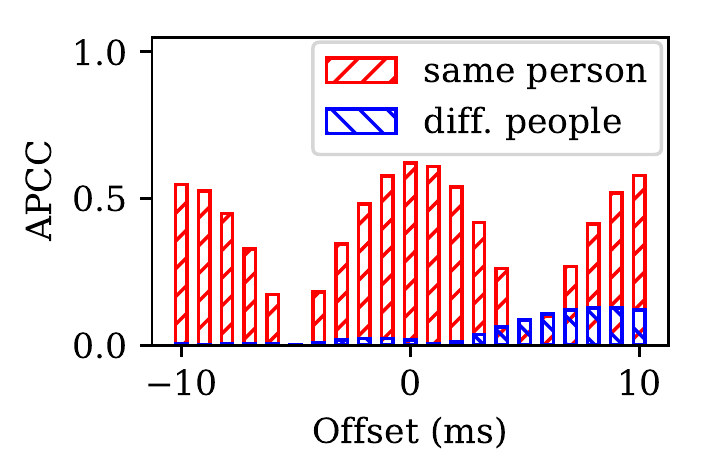}
    \caption{\rev APCC under different clock offsets.}
    \label{fig:clocksync}
  \end{minipage}
  \end{figure}

We use the absolute Pearson correlation coefficient (APCC) to quantify the similarity between two iBEP signals.
{\rev The two bars in the first bar group labeled (a) in Fig.~\ref{fig:pearson} show the APCCs between two iBEP signals collected from the same and different arms on the same person, respectively. The above results suggest that the correlation between the iBEPs is affected by the distance between the sensors. When the two sensors are closer, their iBEPs exhibit higher correlation. This is supportive of our discussion in \sect\ref{subsec:objective}.}

\subsubsection{iBEPs on different bodies}
\label{subsubsec:h2-test}

\begin{figure}
  \centering
  \subfigure[Two nodes in A's right palm.]
  {
    \includegraphics[width=.47\columnwidth]{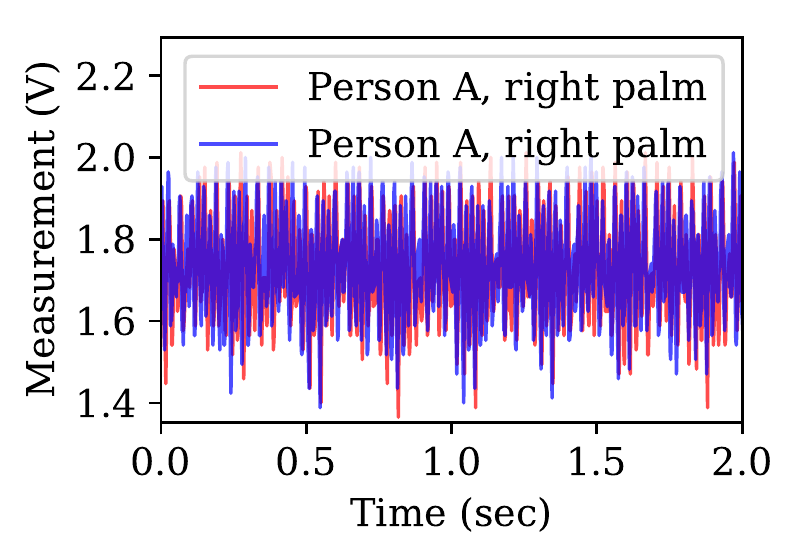}
    \label{fig:sit-silent-same-body}
  }
  \hfill
  \subfigure[Two nodes on two persons.]
  {
    \includegraphics[width=.47\columnwidth]{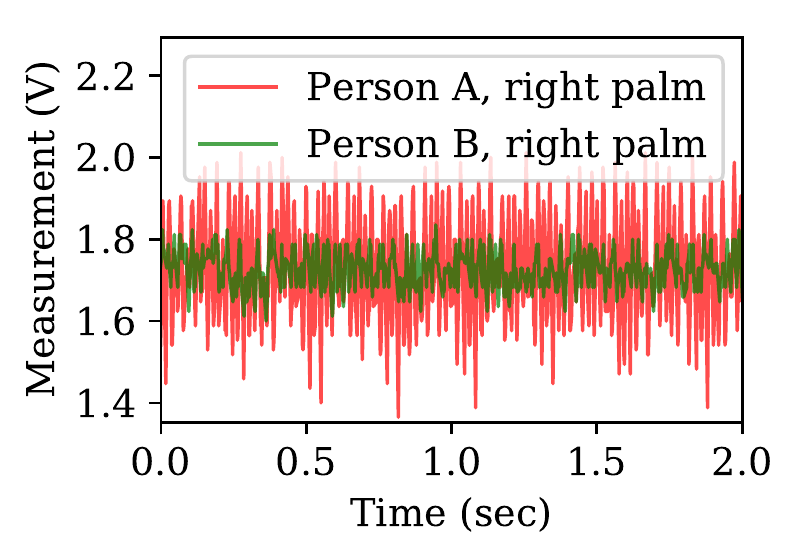}
    \label{fig:sit-silent-diff-body}
  }
  
  \subfigure[Zoomed-in view.]
  {
    \includegraphics[width=\columnwidth]{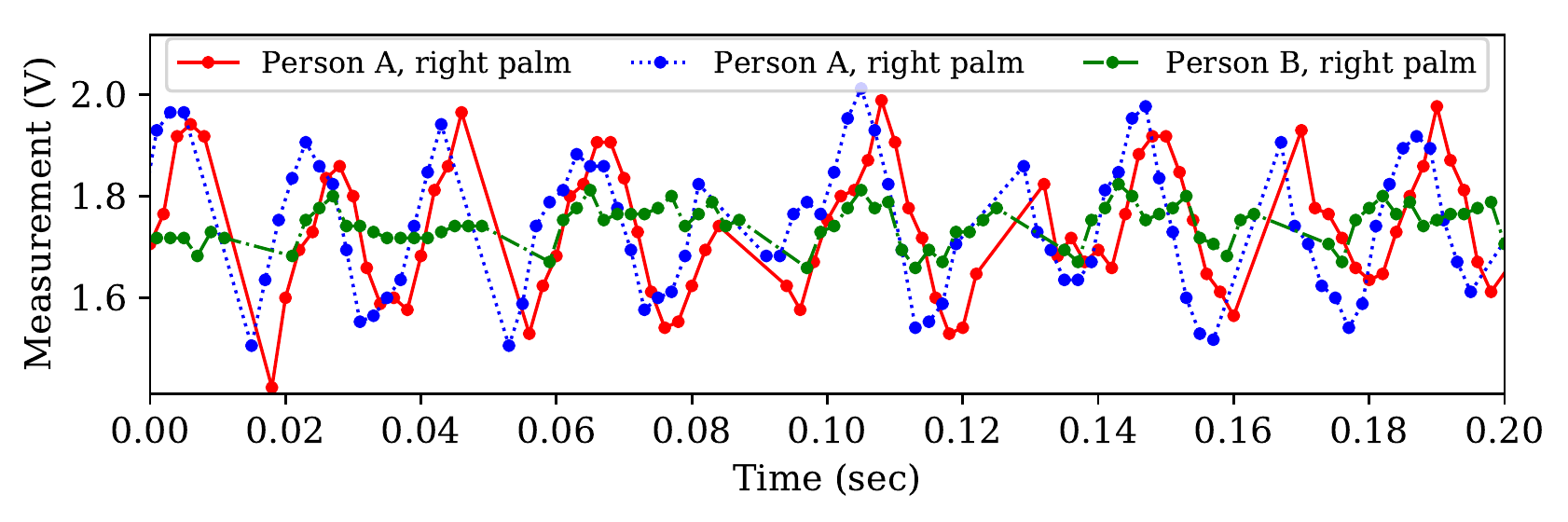}
    \label{fig:zoom-in-sit-silent}
  }
  \caption{iBEPs measured by three sensors on two persons who sit steadily $1\,\text{m}$ apart.}
  \end{figure}

In the first experiment, we place two sensors in the palm of Person~A and another sensor in the palm of Person~B. The two persons sit steadily $1\,\text{m}$ apart. In Fig.~\ref{fig:placements}, the nodes numbered \ding{182}, \ding{183}, and \ding{187} illustrate the placement of the three sensors. Fig.~\ref{fig:sit-silent-same-body} and Fig.~\ref{fig:sit-silent-diff-body} show the iBEPs measured by the two sensors in Person~A's palm and the two persons' palms, respectively. Fig.~\ref{fig:zoom-in-sit-silent} shows the zoomed-in view. From the results, we can see that the iBEP on Person~B is clearly different from that on Person~A, in terms of both signal amplitude and waveform. In contrast, the iBEPs on Person~A are very similar. The two bars in the second bar group labeled (b) in Fig.~\ref{fig:pearson} show the APCCs for the cases shown in Figs.~\ref{fig:sit-silent-same-body} and \ref{fig:sit-silent-diff-body}. Clearly, the iBEPs from the same body exhibit higher correlation than those from different bodies.
{\rev This result is supportive of our discussion in \sect\ref{subsec:objective}.}

\begin{figure}
  \centering
  \subfigure[Two nodes in A's right palm.]
  {
    \includegraphics[width=.47\columnwidth]{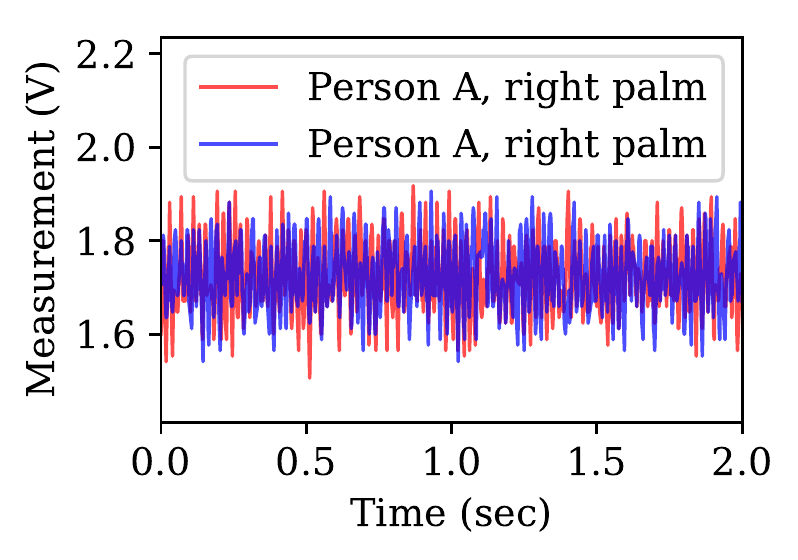}
    \label{fig:moving-same}
  }
  \hfill
  \subfigure[Two nodes on two persons.]
  {
    \includegraphics[width=.47\columnwidth]{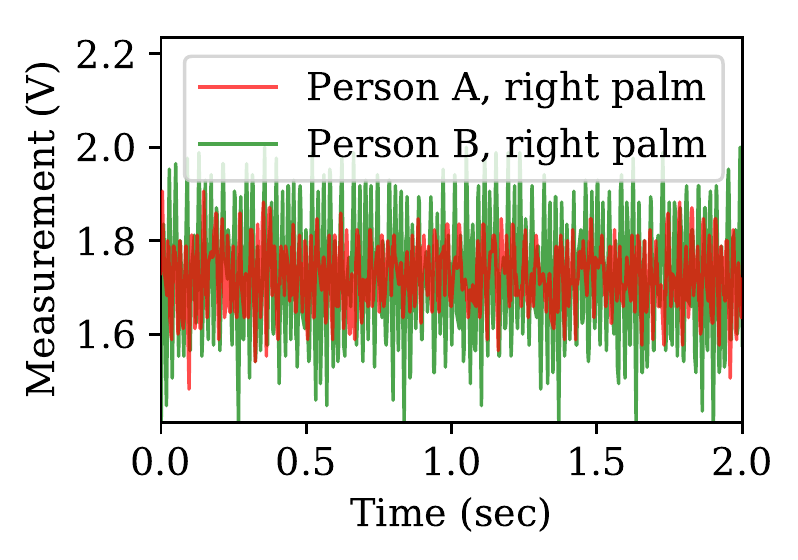}
    \label{fig:moving-diff}
  }
  
  \subfigure[Zoomed-in view.]
  {
    \includegraphics[width=\columnwidth]{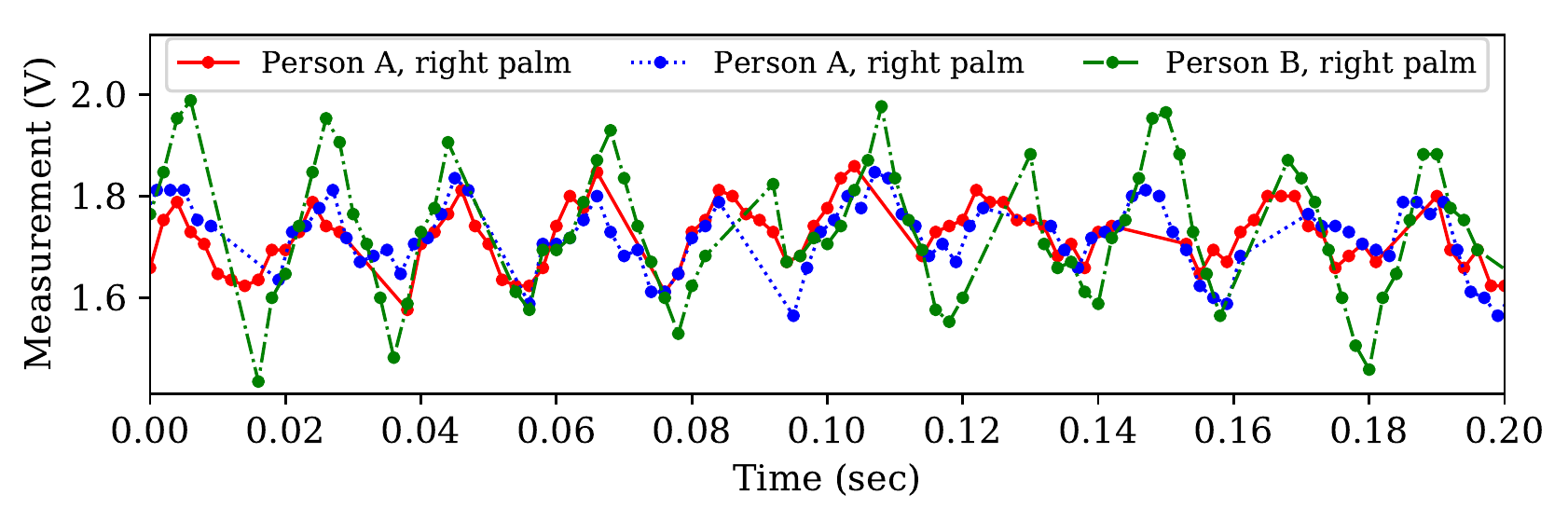}
    \label{fig:moving-zoom-in}
  }
  \caption{iBEPs measured by three sensors on two persons who sit $1\,\text{m}$ apart and perform random hand movements.}
  \label{fig:BEP-gesture}
\end{figure}

In the second experiment, we investigate whether movements will affect the distinctiveness. We ask the two persons to perform some random hand movements. Fig.~\ref{fig:BEP-gesture} and the third bar group labeled (c) in Fig.~\ref{fig:pearson} show the results. We can see that, in the presence of movements, the iBEPs from the same body still exhibit higher correlation than those from different bodies.

In the above experiments, the clocks of the sensors are tightly synchronized using FTSP that uses MAC-layer timestamping to achieve microsecond-level synchronization accuracy. Platforms without MAC-layer timestamping can achieve millisecond-level synchronization accuracy \cite{li2012flight}.
{\rev We now assess the impact of a clock synchronization error of up to $10\,\text{ms}$ on the APCC. As our collected iBEP signals are tightly synchronized, we simulate the clock synchronization error by offseting an input iBEP signal. Fig.~\ref{fig:clocksync} shows the APCC under different simulated clock offsets among the signals in Fig.~\ref{fig:BEP-gesture}. We can see that, in the presence of clock synchronization error, the APCC for the signals from the same person is generally higher than that for different persons. Moreover, when the synchronization error is around $-5\,\text{ms}$ or $5\,\text{ms}$, the APCCs are nearly zero. This is because the two signals have a phase difference of $90\text{\textdegree}$, resulting in near-zero correlations. Our earlier study \cite{yan2017application} shows that by using iBEP, wearables on the same person or two nearby persons can maintain the synchronization errors below $3\,\text{ms}$. Such synchronization errors will not subvert the APCC as an effective similarity metric.}

\section{Same-Body Contact Detection}
\label{sec:approach}

{\rev
From \sect\ref{subsec:measurement-results}, iBEP is promising for touch-to-access device authentication.
In this section, we present the design of the same-body contact detection algorithm (\sect\ref{subsec:approach-algo}) and discuss a {\em mimicry attack} that aims at subverting the algorithm (\sect\ref{subsec:security1}).
}

\subsection{Detection Algorithm}
\label{subsec:approach-algo}

Before TouchAuth detects the same-body contact, it checks the iBEP signal strength. Specifically, if the standard deviation of either $s(t)$ or $s'(t)$ is below a predefined threshold, TouchAuth rejects the authentication request without performing same-body contact detection. This ensures that the detection is made based on meaningful iBEP signals. From our offline tests, a standard deviation threshold of $0.06\,\text{V}$ is a good setting for the Z1 platform. Similar offline tests can be performed for other platforms. In what follows, we present the same-body contact detection algorithm. The detection performance will be evaluated in \sect\ref{sec:eval}.

\subsubsection{Similarity-based detector}
\label{subsubsec:similarity}

The detector compares a similarity score between $s(t)$ and $s'(t)$, $\forall t \in [t_1, t_2]$, with a threshold denoted by $\eta$. If the similarity score is larger than $\eta$, TouchAuth accepts the authenticatee; otherwise, it rejects the authenticatee. We adopt the reciprocal of the root mean square error (RMSE) and the absolute Pearson correlation coefficient (APCC) as our similarity metrics. The RMSE is a variant of the Euclidean distance which has been used as a dissimilarity metric by physiological sensing approaches \cite{poon2006novel}. The Pearson correlation coefficient measures the linear correlation between two variables. As shown in Fig.~\ref{fig:same-body-diff-arms-all}, the iBEP signals collected from the same arm have a phase shift of $180^{\circ}$, resulting in a Pearson correlation of about $-1$. However, the authenticatee on the same arm as the authenticator may be accepted. This motivates us to use the APCC as the similarity metric that ranges from $0$ to $1$, with $0$ and $1$ representing the lowest and the highest similarity values, respectively. In the rest of this paper, the TouchAuth based on the RMSE and APCC is called {\em RMSE-TouchAuth} and {\em APCC-TouchAuth}, respectively.

Dynamic time warping distance (DTWD) is also a widely adopted dissimilarity metric that can address time-varying phase shift. From our experiments, it may wrongly help the invalid authenticatee who is spatially close to the authenticator. Thus, we do not adopt DTWD.

\subsubsection{Assessment metrics}
\label{subsubsec:parameter-configuration}

This paper uses the false acceptance rate (FAR or simply $\alpha$) and the true acceptance rate (TAR or simply $\beta$) as the main detection performance metrics. The $\alpha$ and $\beta$ are the probabilities that an invalid or valid authenticatee is wrongly or correctly accepted, respectively. The detection threshold $\eta$ and the signal length $\ell$ are two important parameters. The receiver operating characteristic (ROC) curve of $\beta$ versus $\alpha$ by varying $\eta$ depicts fully the performance of a detector under a certain $\ell$. The signal length $\ell$ characterizes the sensing time needed by the authentication process. In this paper, we use the ROC curves to compare the detection performance of various detectors. In practice, the settings of $\eta$ and $\ell$ can follow the Neyman-Pearson lemma to enforce an upper bound for $\alpha$. A stringent $\alpha$ is often required by authentication. For instance, with $\alpha=1\%$, an invalid authenticatee needs to repeat the authentication process 100 times on average to be successful, which is frustrating if some after-rejection freeze time is enforced. Moreover, an authenticatee device can be banned if it is continuously rejected for many times.

\begin{figure}
  \vspace{-0.7em}
  \subfigure[FAR \& FRR vs. threshold $\eta$.]
  {
    \includegraphics[width=.47\columnwidth]{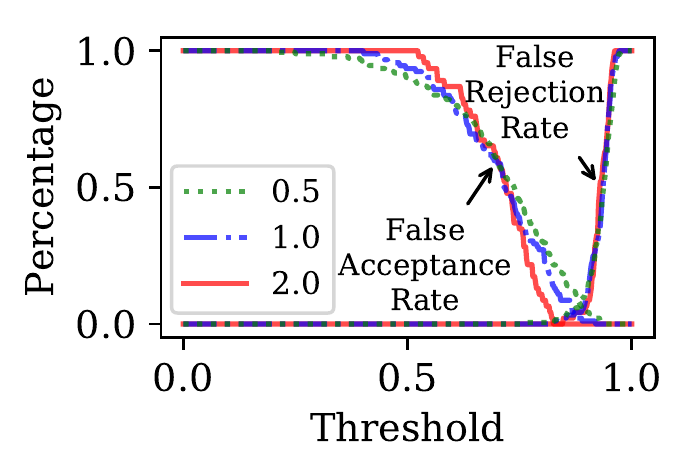}
    \label{fig:apcc-threshold}
  }
  \hfill
  \subfigure[ROC curves.]
  {
    \includegraphics[width=.46\columnwidth]{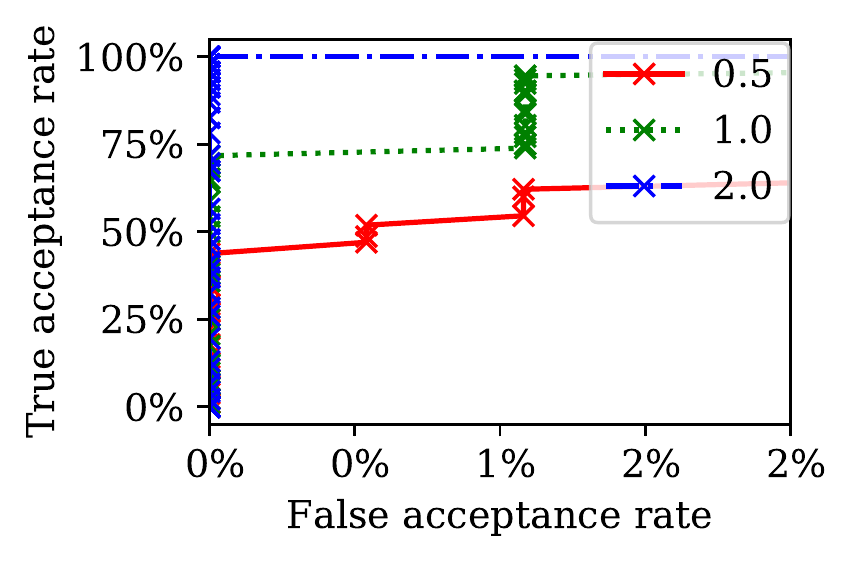}
    \label{fig:apcc-roc}
  }
  \caption{Detection performance of APCC-TouchAuth when $\ell$ is $0.5\,\text{s}$, $1\,\text{s}$, and $2\,\text{s}$, respectively.}
  \label{fig:apcc-touch}
\end{figure}

Fig.~\ref{fig:apcc-touch} shows the detection performance of APCC-TouchAuth assessed by using the data shown in Fig.~\ref{fig:BEP-gesture}. Fig.~\ref{fig:apcc-threshold} shows the $\alpha$ and the false rejection rate (FRR) versus the detection threshold $\eta$ when $\ell$ is $0.5\,\text{s}$, $1\,\text{s}$, and $2\,\text{s}$, respectively. Note that $\text{FRR}=1-\beta$. Fig.~\ref{fig:apcc-roc} shows the ROC curves when $\alpha$ is from 0 to 2\%. Note that the $\alpha$ and $\beta$ values of each point on the ROC are measured based on 500 tests. From the figure, we can see that when $\ell=2\,\text{s}$, APCC-TouchAuth achieves a $\beta$ value of 100\% (i.e., correctly accepts all 500 tests when the authenticatee is valid) while keeping $\alpha=0\%$ (i.e., correctly rejects all 500 tests when the authenticatee is invalid). This suggests that APCC-TouchAuth can achieve a very high detection accuracy. From Fig.~\ref{fig:apcc-roc}, the ROC curve under a smaller $\ell$ setting becomes lower, suggesting lower detection accuracy. \sect\ref{sec:eval} will extensively evaluate the detection performance of TouchAuth under a wider range of settings among a larger group of users.

In addition to the ROC that characterizes detection performance, we also use the {\em signal-to-difference ratio} (SDR) to assess the quality of iBEP sensing. Specifically, let $P[x(t)]$ denote the average power of a signal $x(t)$. Ideally, if the authenticator and the valid authenticatee are very close to each other on the same human body, their iBEP signals $s(t)$ and $s'(t)$ should be very similar. Thus, we define the SDR in decibel as $\mathrm{SDR} = 10 \log_{10} \frac{P[s(t)]}{P[s(t) - s'(t)]}$ dB. A high SDR suggests high-quality iBEP sensing.

\subsection{Mimicry Attack}
\label{subsec:security1}

We now discuss a {\em mimicry attack} that attempts to obtain the authenticator's $s(t)$. Due to the complex spatial distribution of the indoor ambient EF, it is generally difficult for the attacker to estimate the authenticator's $s(t)$. In this attack, the attacker wearing an iBEP sensor mimics the body movements of the victim user wearing the authenticator. To be effective, the mimicry attacker should stay as close as possible to the victim user to sense the same/similar ambient EF. Thus, it is unrealistic in practice, because the strange mimic behavior in proximity can be easily discerned by the user. {\rev Note that this attack is beyond the threat model defined in \sect\ref{subsec:threat-model} that concerns the security of data communications between the authenticator and the authenticatee. Thus, our approach described in \sect\ref{subsec:approach-overview}, which is based on the secure protocol H2H, does not guarantee security against this mimicry attack. In \sect\ref{sec:eval}, we will show the ineffectiveness of this attack experimentally.}

\section{Evaluation}
\label{sec:eval}

We conduct a set of experiments to evaluate TouchAuth's same-body contact detection performance under a wide range of settings including different wearers, various indoor environments, multiple possible interfering sources, device proximity, {\rev skin moisture}, and heterogeneous devices.

\subsection{Performance across Different Wearers}
\label{subsubsec:diff-persons}

We collect a set of data involving a wearer $\mathcal{R}$ and 12 other wearers $\mathcal{P}_1, \mathcal{P}_2, \ldots, \mathcal{P}_{12}$. The experiments are conducted in a computer science lab. In the $i$th experiment ($i=1, \ldots, 12$), $\mathcal{R}$ holds an authenticator device and a valid authenticatee device in his palm, whereas $\mathcal{P}_i$ holds an invalid {\rev authenticatee} device in his palm. Thus, in this set of experiments, we evaluate the detection performance of TouchAuth for a certain user with a valid authenticatee against different users with invalid authenticatees.
In each experiment, $\mathcal{R}$ and $\mathcal{P}_i$, which are about $0.5\,\text{m}$ apart, are allowed to perform some uncoordinated and random hand movements. The data collection of each experiment lasts for two minutes. We measure the detection performance of APCC-TouchAuth and RMSE-TouchAuth as follows. Let $N_{L}$, or $N_{I}$, denote the total number of tests between the authenticator and the valid authenticatee, or between the authenticator and the invalid authenticatee. Accordingly, let $N_{TA}$ and $N_{FA}$ denote the total numbers of true acceptances and false acceptances, respectively. The $\beta$ and $\alpha$ are measured by $N_{TA}/N_{L}$ and $N_{FA}/N_{I}$, respectively.

\begin{figure}
  \centering
  \includegraphics[width=\columnwidth]{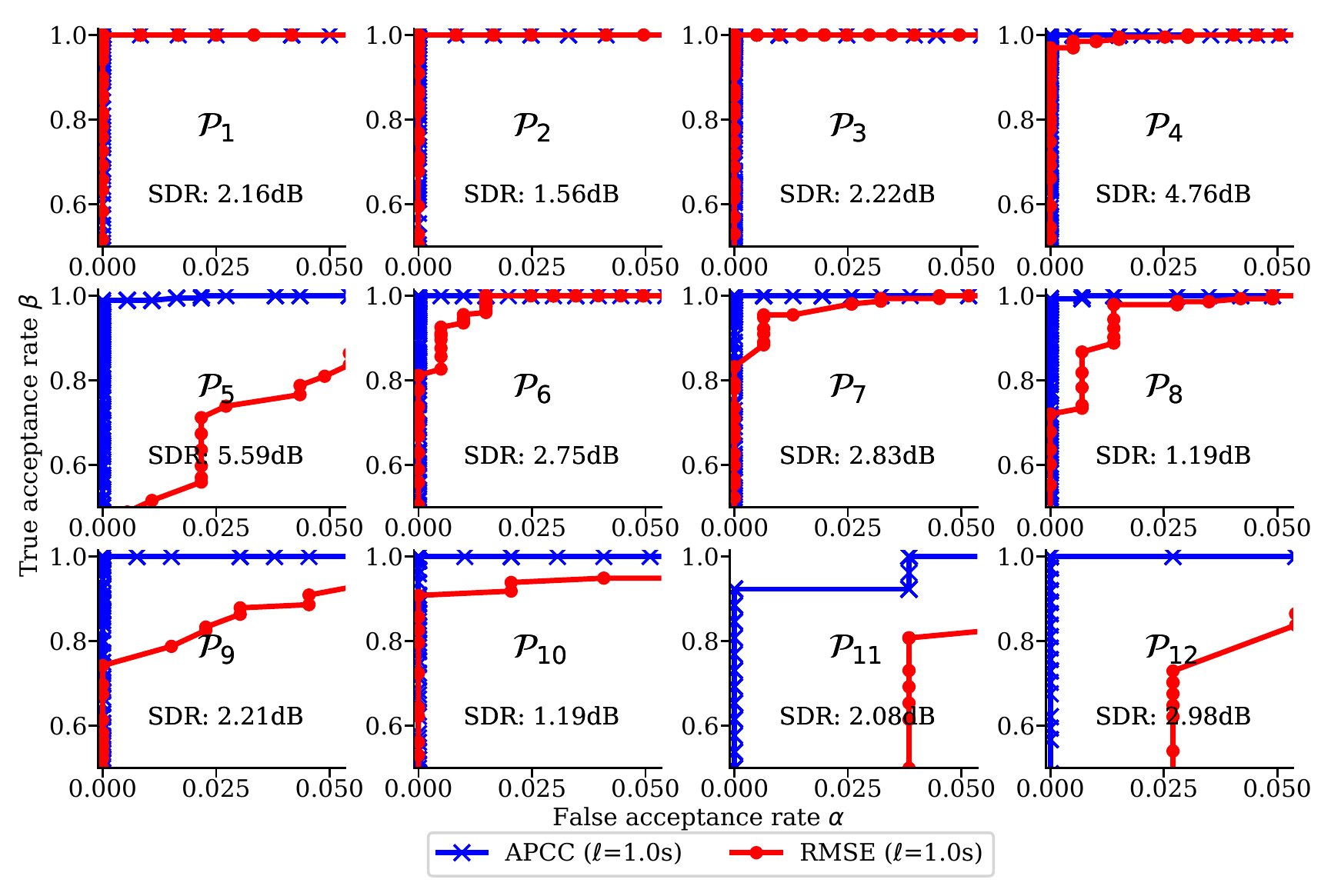}
  \caption{ROCs for 12 different wearers with the invalid authenticatee device. The $x$-axis and $y$-axis of each subfigure are $\alpha$ and $\beta$, respectively.}
  \label{fig:eva-roc-diff-ppl}
  \end{figure}

Fig.~\ref{fig:eva-roc-diff-ppl} shows the APCC- and RMSE-TouchAuth's ROC curves for different wearers with the invalid authenticatee device when the signal length $\ell$ is $1\,\text{s}$. Different data points on an ROC represent the results under different detection threshold $\eta$. The SDR assessed using the authenticator's and the valid authenticatee's iBEP signals in each experiment is included in the corresponding subfigure. We can see that across different wearers with the invalid authenticatee, APCC-TouchAuth is comparable or superior to RMSE-TouchAuth in terms of the detection performance. This is because that APCC inherently captures the correlation between the iBEP signals on the same moving hand. In contrast, as the RMSE captures sample-wise differences between two signals, two uncorrelated signals with similarly small amplitudes can give a small RMSE value, leading to a false acceptance. Note that the RMSE has been adopted as a dissimilarity metric for physiological sensing \cite{poon2006novel}. However, it is ill-suited for iBEP sensing because the iBEP signal amplitude has a large dynamic range depending on the ambient EF's gradient. This is different from physiological signals that often have stable ranges of signal amplitude. From Fig.~\ref{fig:eva-roc-diff-ppl}, we can see that APCC-TouchAuth achieves a high $\beta$ value (100\%) subject to an $\alpha$ upper bound of 1\%, except for the wearer $\mathcal{P}_{11}$. For $\mathcal{P}_{11}$, APCC-TouchAuth achieves a $\beta$ value of 100\% subject to an $\alpha$ upper bound of 4\%.

\begin{figure}
  \centering
  \includegraphics[width=\columnwidth]{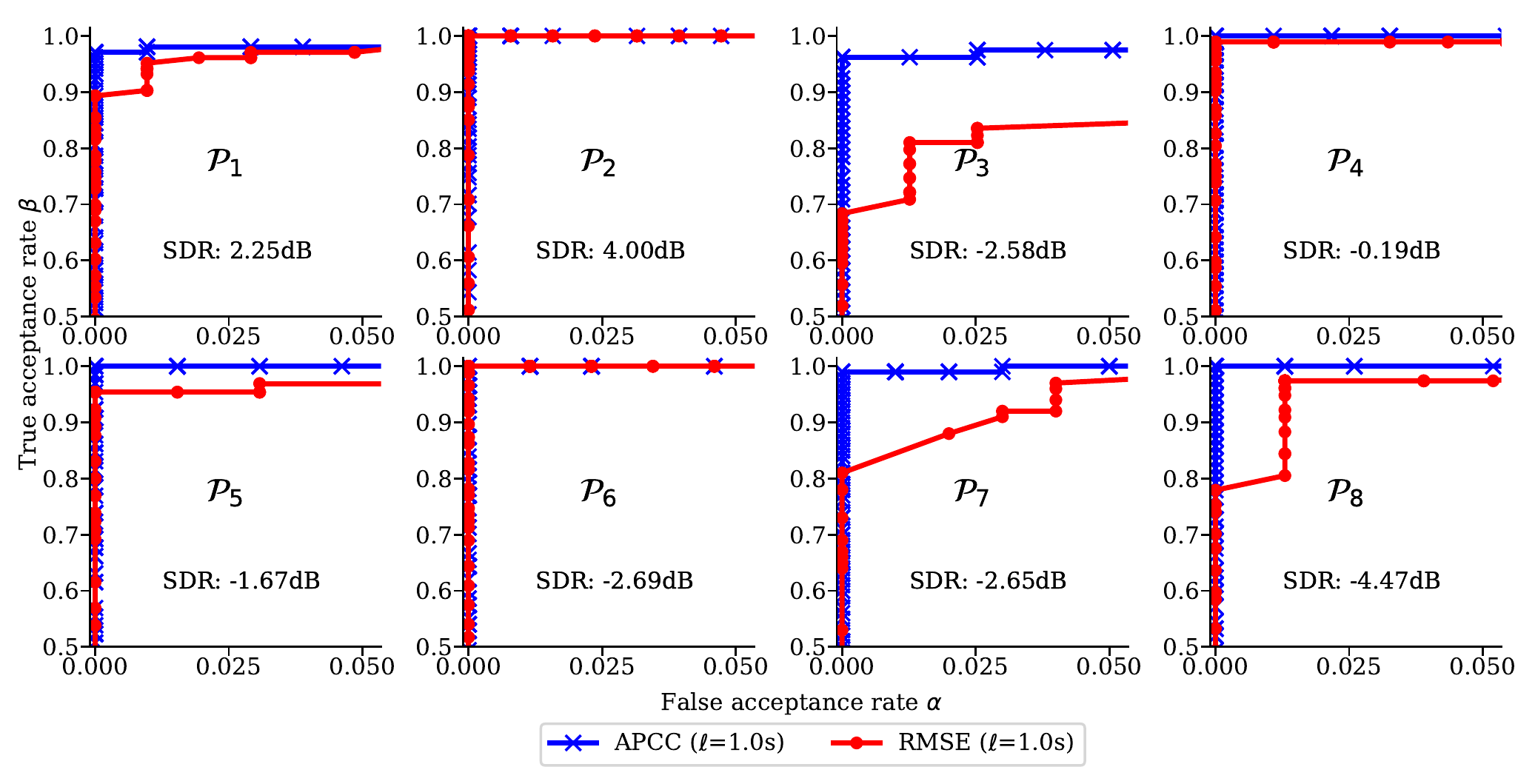}
  \caption{ROCs for 8 different wearers with the valid authenticatee device. The $x$-axis and $y$-axis of each subfigure are $\alpha$ and $\beta$, respectively.}
  \label{fig:eva-roc-diff-ppl2}
  \end{figure}

\begin{figure}
  \centering
    \includegraphics[width=.85\columnwidth]{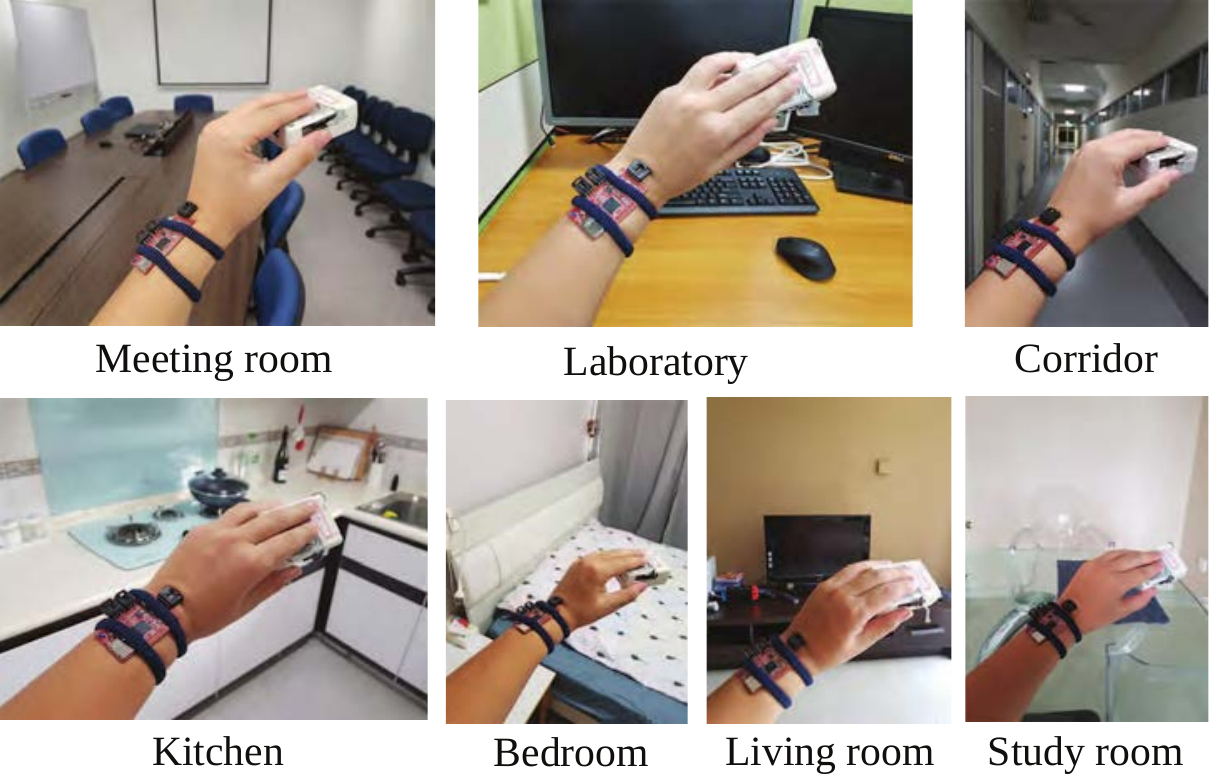}
    \caption{Experiments in indoor environments.}
    \label{fig:indoor-environments}
\end{figure}

\begin{figure}
  \centering
    \includegraphics[width=\columnwidth]{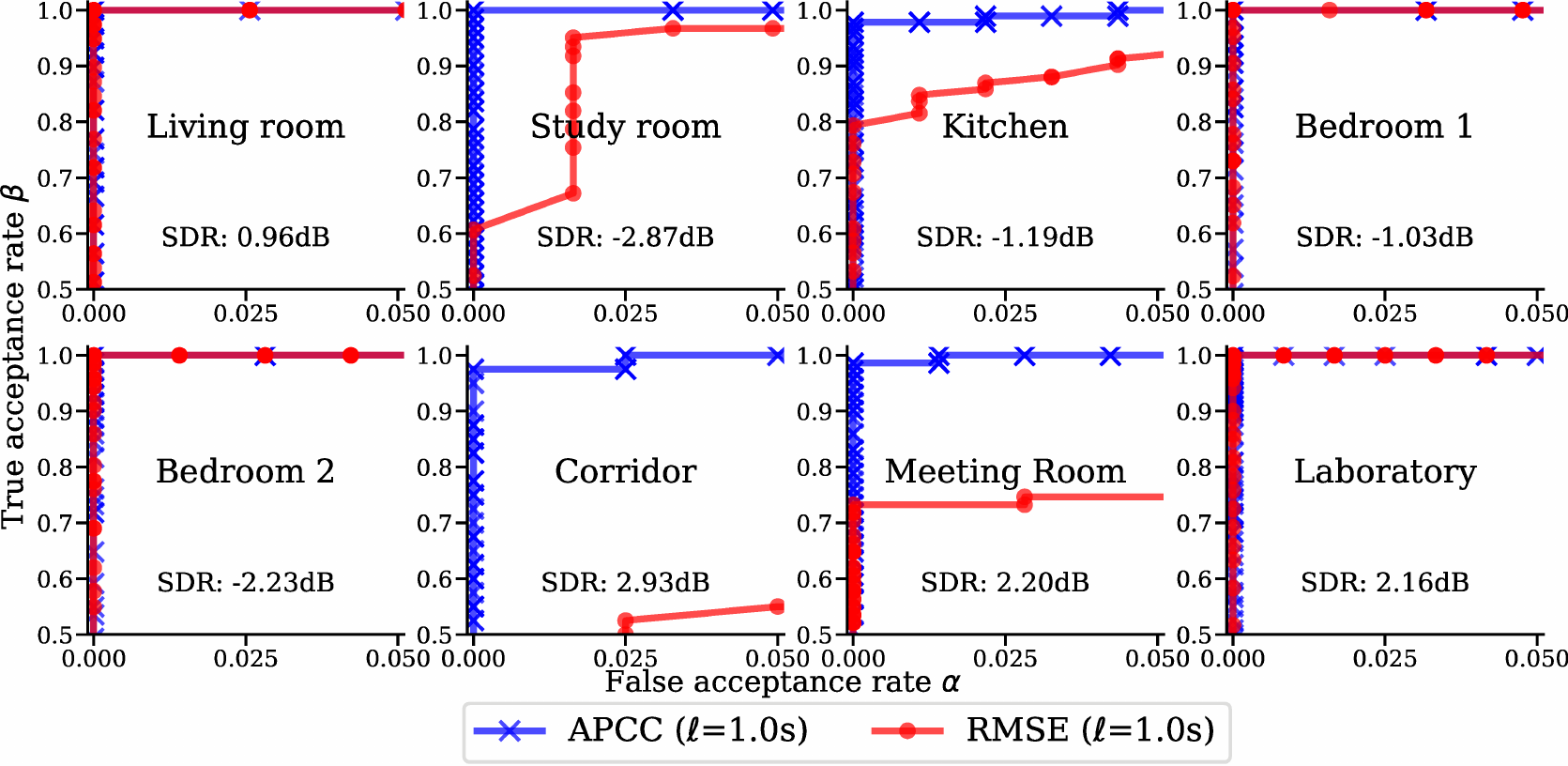}
    \caption{ROCs in various environments.}
    \label{fig:eva-roc-diff-loc}
    \vspace{-1em}
\end{figure} 

We collect another set of data, where $\mathcal{R}$ wears an invalid authenticatee and $\mathcal{P}_i$ holds an authenticator and a valid authenticatee. Thus, this set of experiments evaluate the detection performance of TouchAuth for different users wearing the valid authenticatee against a certain user wearing the invalid authenticatee. Fig.~\ref{fig:eva-roc-diff-ppl2} shows the ROCs for eight different wearers with the valid authenticatee. Similar to the results in Fig.~\ref{fig:eva-roc-diff-ppl}, APCC-TouchAuth achieves high-profile ROCs and outperforms RMSE-TouchAuth. The results in Figs.~\ref{fig:eva-roc-diff-ppl} and \ref{fig:eva-roc-diff-ppl2} show that the detection performance of TouchAuth is not wearer-specific.

\subsection{Various Indoor Environments}
\label{subsec:indoor-environments}

Two wearers conduct experiments in eight different indoor environments as shown in Fig.~\ref{fig:indoor-environments}, which include a living room, a study room, a kitchen, two bedrooms, a corridor, a meeting room, and an open area of a lab. One wearer carries the authenticator and a valid authenticatee and the other carries the invalid authenticatee. Fig.~\ref{fig:eva-roc-diff-loc} shows the snapshots of some environments and the measured SDRs and ROCs in the eight environments. We can see that in certain environments, the RMSE-TouchAuth performs poorly. Investigation on the raw iBEP signals shows that in these environments, the iBEP signals of the authenticator and the invalid authenticatee have similar amplitudes. In all the eight environments, APCC-TouchAuth achieves high $\beta$ values ($\ge 97\%$) subject to an $\alpha$ upper bound of 1\%. If the $\alpha$ upper bound is relaxed to 4\%, the $\beta$ value of 100\% can be achieved.

\subsection{Various Possible Interfering Sources}
\label{subsec:interfering}
\begin{figure}
  \centering
  \includegraphics[width=\columnwidth]{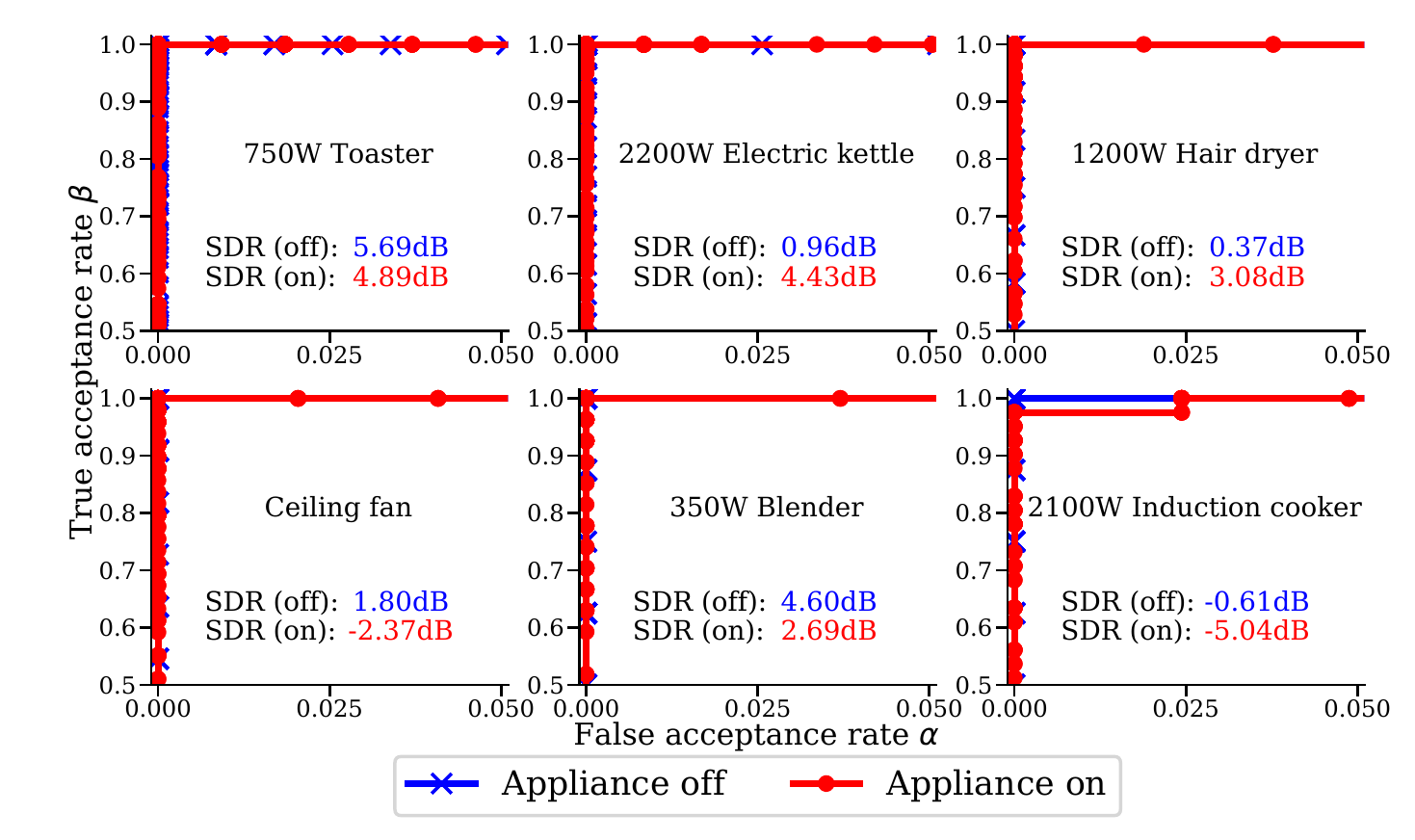}
  \caption{ROCs with various nearby appliances.}
  \label{fig:eva-roc-diff-app}
\vspace{-1em}
\end{figure}

From our discussion in \sect\ref{sec:basis} and the measurement results in \sect\ref{subsubsec:insensitivity-mf}, the iBEP measurement is mainly caused by the ambient EF. The MFs generated by the operating currents of electric appliances will have little impact on the iBEP sensing. However, some appliances, especially those based on motors and high-frequency switched-mode power, may generate interference to the iBEP sensing. This is because that unlike the $50\,\text{Hz}$ current-induced MF that generates little/no EF, the high-frequency currents caused by the frictions between the motor's brush and stator as well as the switched-mode power may generate propagating electromagnetic waves. {\rev As a result, the EFs generated by the appliances and powerlines may weaken each other, making the overall EF weaker.} Thus, we conduct a set of experiments with various home appliances including toaster, electric kettle, hair dryer, ceiling fan, blender, and induction cooker. Specifically, two wearers, one with a valid authenticatee and the other with an invalid authenticatee, stand close to a certain appliance to collect iBEP traces. Fig.~\ref{fig:eva-roc-diff-app} shows the SDR and the APCC-TouchAuth's ROCs for various appliances when the appliance is on and off. We can see that, for a certain appliance, the SDR may increase or decrease when the appliance is switched on. This is because that the interference from the appliance may be constructive or destructive to the EF generated by the building's power cabling. The operating status of the induction cooker causes the largest SDR change of more than $5\,\text{dB}$. This is due to the high-frequency switched-mode current in the cooker's internal inductor. As a result, the ROC drops slightly when the induction cooker is switched on. However, APCC-TouchAuth still achieves a high $\beta$ value (100\%) subject to an $\alpha$ upper bound of $2.5\%$.

\begin{figure}
  \includegraphics[width=\columnwidth]{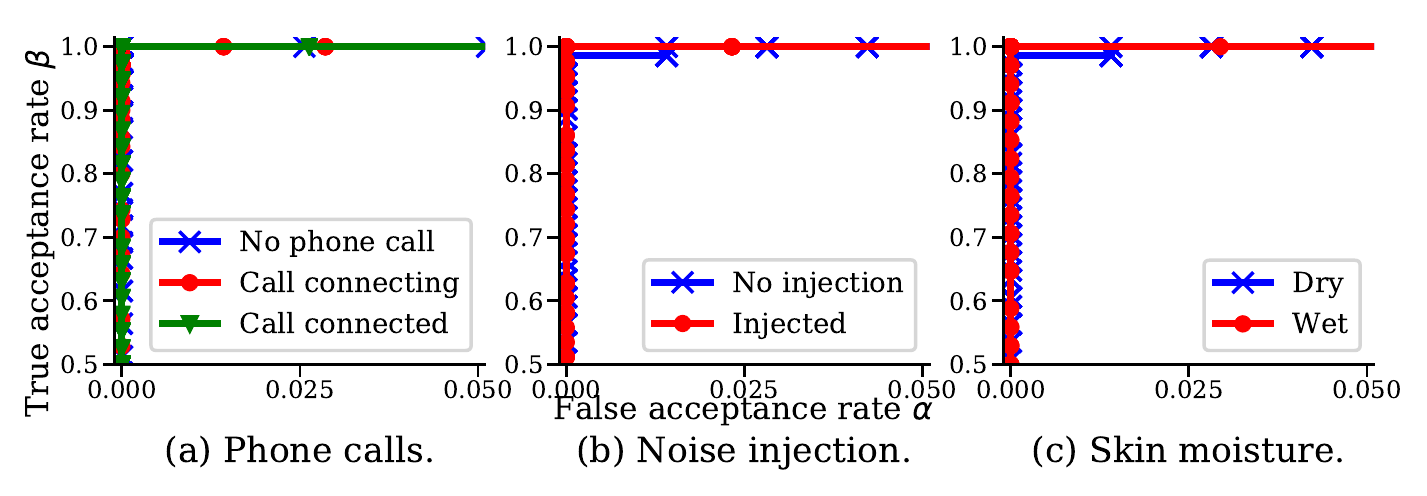}
  \caption{ROCs with interference and skin moisture.}
  \label{fig:eva-roc-interfere}
\end{figure}

The signaling phase of a cell phone call often interferes with audio systems because of the intermittent wireless power pulses. Thus, we also evaluate the impact of cell phone calls on APCC-TouchAuth. In the experiments, the wearer holds a smartphone, a valid authenticatee, and the authenticator in one palm. Another wearer holding an invalid authenticatee stands $0.5\,\text{m}$ away. Fig.~\ref{fig:eva-roc-interfere}(a) shows the ROCs at different phases of a phone call. We can see that the phone call does not affect the detection performance of APCC-TouchAuth.

Secondly, we use a circuit seeker (Greenlee CS-8000) that is capable of up to 4 miles circuit tracing \cite{circuit-seeker} to inject noises into the power network serving the lab in which we conduct experiments. The injector of CS-8000 is plugged into a power outlet, injecting a $15\,\text{kHz}$ signal into the power network; the seeker can detect the $15\,\text{kHz}$ electromagnetic emanation from the powerlines. We conduct experiments in proximity of a powerline close to the injector. Fig.~\ref{fig:eva-roc-interfere}(b) shows the ROCs when the injector is in operation or not. We can see that the noise injection does not affect TouchAuth.

{\rev 
Lastly, we evaluate the impact of the skin moisture conditions on TouchAuth. We conduct two experiments, in which the user holds the authenticator using a wet hand. He also holds a valid authenticatee. Another user stands $0.5\,\text{m}$ away holding an invalid authenticatee. Fig.~\ref{fig:eva-roc-interfere}(c) shows the ROCs for dry and wet skin moisture conditions. We can see that the skin moisture has little impact on the performance of TouchAuth.
}

\subsection{Impact of Signal Length $\ell$}
\label{subsubsec:decision-fusion}

\begin{figure}
  \centering
  \begin{minipage}[t]{.47\columnwidth}
    \vspace{0pt}
    \includegraphics[width=\textwidth]{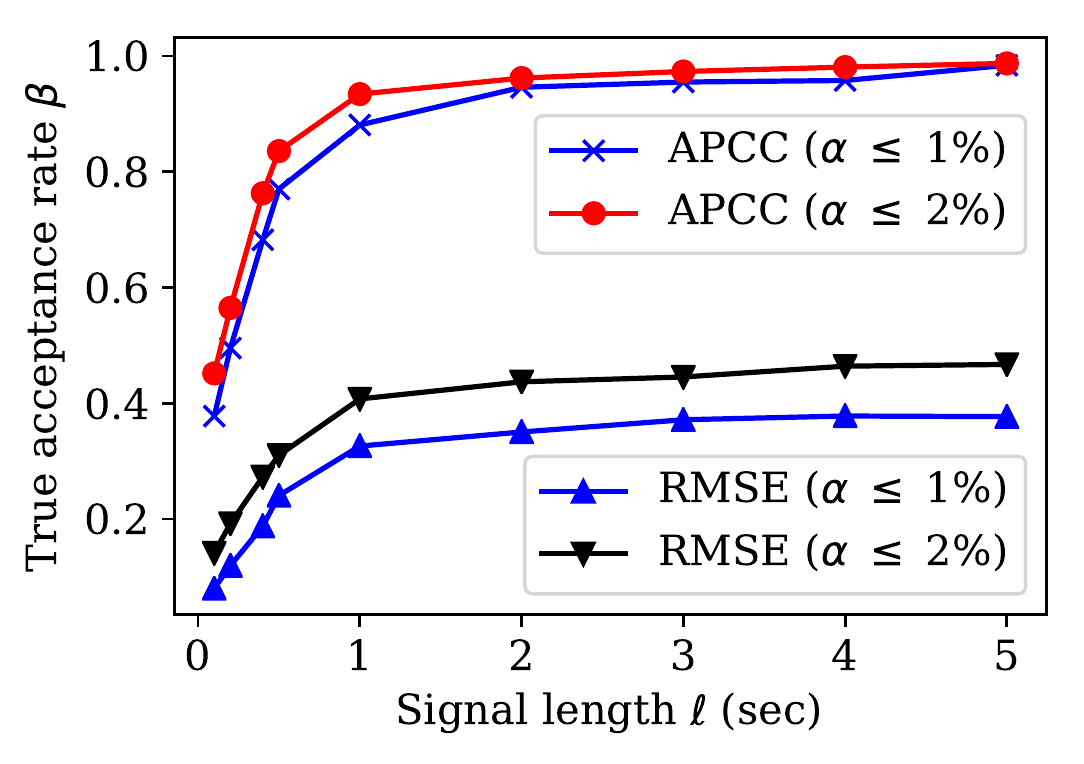}
    
    \caption{$\beta$ versus $\ell$.}
    \label{fig:eva-window-len}
  \end{minipage}
  \begin{minipage}[t]{.52\columnwidth}
    \vspace{-0.2em}
    \centering
    \includegraphics[width=.93\textwidth]{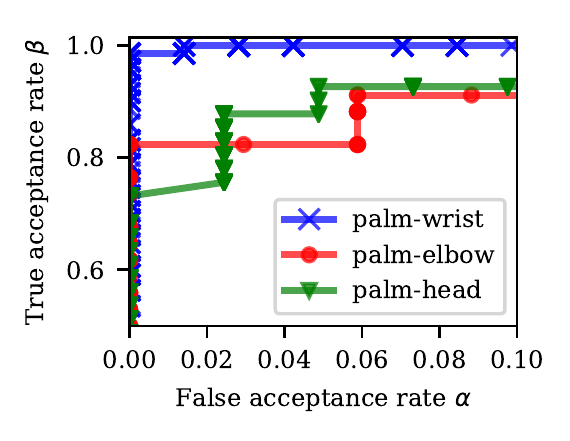}
    \vspace{-0.8em}
    \caption{Sensor proximity.}
    \label{fig:near-field}
  \end{minipage}
  \vspace{-1em}
\end{figure}

We evaluate the impact of the signal length $\ell$ on the detection performance of TouchAuth. We combine the data collected from 12 different wearers in \sect\ref{subsubsec:diff-persons} into a single dataset. Based on the combined dataset, Fig.~\ref{fig:eva-window-len} shows the $\beta$ achieved by APCC-TouchAuth and RMSE-TouchAuth versus $\ell$ when $\alpha \le 1\%$ or $\alpha \le 2\%$. APCC-TouchAuth's $\beta$ increases sharply when $\ell \le 1\,\text{s}$. When $\ell > 1\,\text{s}$, its $\beta$ increases with $\ell$ slowly. This suggests that a setting of $\ell=1\,\text{s}$ well balances the detection performance and sensing time.
The $\beta$-$\ell$ curves for RMSE-TouchAuth exhibits a similar pattern. Moreover, consistent with the results in \sect\ref{subsubsec:diff-persons} and \ref{subsec:indoor-environments}, RMSE-TouchAuth is inferior to APCC-TouchAuth.

\subsection{TouchAuth Devices' Proximity}
\label{subsec:near-field}

In the previous subsections, the authenticator and the valid authenticatee are in the same palm. In this set of experiments, they are placed at different locations on the user's body. Fig.~\ref{fig:near-field} shows APCC-TouchAuth's ROC curves. When the two devices are on the palm and the wrist of the same hand, respectively, a high-profile ROC is achieved. When the two devices are on (i) the right palm and the right elbow, respectively, or (ii) the right palm and the head, respectively, the detection performance is degraded. This shows that TouchAuth is applicable to the example use scenarios discussed in \sect\ref{sec:intro} where the two devices are in proximity on the same body. In \sect\ref{sec:limit}, we will further discuss the impact of the proximity requirement on the usability of TouchAuth.

\subsection{Mimicry Attack}

\begin{figure} 
  \centering  
  \begin{minipage}[t]{.49\columnwidth}
    \includegraphics[width=\textwidth]{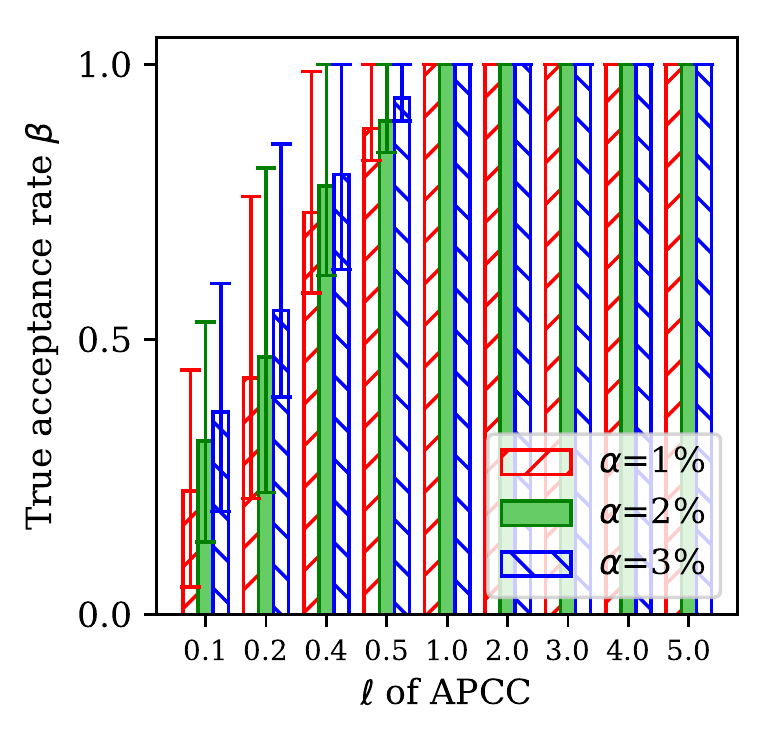}
    \caption{$\beta$ vs. $\ell$ under mimicry attack.}
    \label{fig:mimic}
  \end{minipage}
  \hfill
  \begin{minipage}[t]{.49\columnwidth}
    \includegraphics[width=\textwidth]{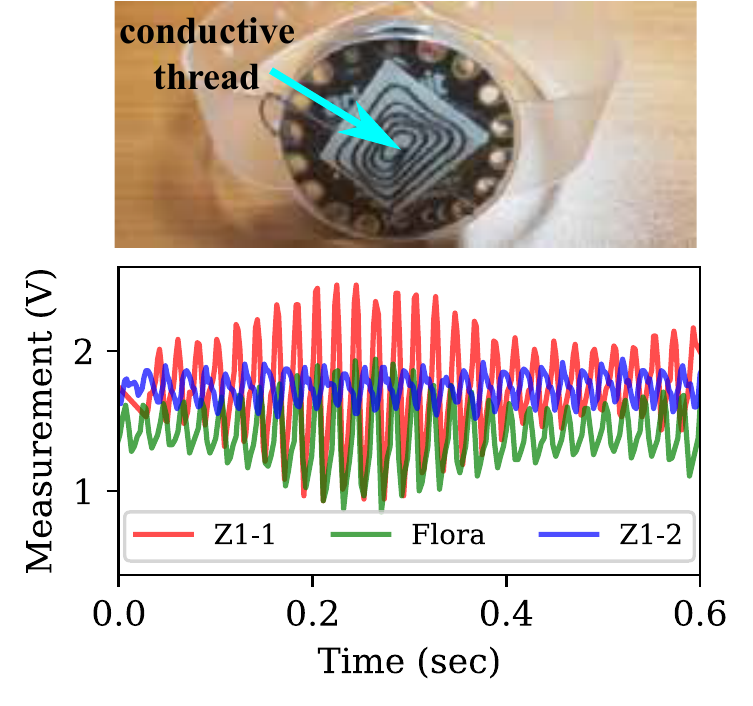}
    \caption{A Flora as the valid authenticatee.}
    \label{fig:flora}
    
  \end{minipage}
  \end{figure}

We follow the data collection methodology described in \sect\ref{subsubsec:diff-persons} to collect another dataset in the lab, except that each wearer $\mathcal{P}_i$ with the invalid authenticatee mimics the hand movements of the wearer $\mathcal{R}$ with the authenticator and the valid authenticatee. The $\mathcal{R}$ performs simple and repeated hand movements,
such that $\mathcal{P}_i$ can follow easily. The distance between $\mathcal{R}$ and $\mathcal{P}_i$ is about $0.5\,\text{m}$. Fig.~\ref{fig:mimic} shows the APCC-TouchAuth's $\beta$ versus $\ell$ subject to various $\alpha$ upper bounds. The error bars show the minimum, maximum, and mean of the $\beta$ values among different $\mathcal{R}$-$\mathcal{P}_i$ pairs in the dataset. Compared with Fig.~\ref{fig:eva-window-len}, when $\ell$ is small (e.g., $0.1\,\text{s}$), the mimicry attack degrades APCC-TouchAuth's detection performance. However, the attack impact can be fully mitigated by adopting a larger $\ell$ setting (e.g., $\ell=1\,\text{s}$).

\subsection{Heterogeneous Devices}
\label{subsec:hetero}

In this set of experiments, the authenticator and the invalid authenticatee are based on Z1 motes (denoted by Z1-1 and Z1-2); the valid authenticatee is based on an Adafruit's Flora \cite{flora}, an Arduino-based wearable platform. The top part of Fig.~\ref{fig:flora} shows a Flora-based TouchAuth prototype device with a 3D-printed insulating wristband and a conductive thread creating the body contact. We use a laptop computer to relay the communications between the Bluetooth-based Flora and the Zigbee-based Z1. The bottom part of Fig.~\ref{fig:flora} shows the zoomed-in view of the signals captured by the three devices. We can see that although the Z1 and Flora have different direct current lines, the Z1-1 authenticator and the Flora authenticatee are highly correlated. Based on this setup, with $\ell=1\,\text{s}$, APCC-TouchAuth achieves a $\beta$ value of 100\% subject to an $\alpha$ upper bound of 1\%. This result shows that TouchAuth can be applied on heterogeneous devices.

\section{Limitations and Discussions}
\label{sec:limit}

This section discusses limitations and the applicable scope of TouchAuth.

\vspace{.5em}
\noindent {\bf Applicability to outdoors:}
The outdoor naturally occurring EF is too weak to be exploited by TouchAuth. Thus, TouchAuth is not applicable outdoors, where it will reject all authentication requests due to too weak iBEP signal strength (cf.~\sect\ref{subsec:approach-algo}). As most smart objects are indoors and we spend most of our time indoors (e.g., 87\% on average for Americans \cite{klepeis2001national}), TouchAuth gives a satisfactory availability.
Note that the wide availability of the iBEP signals in indoor environments has been shown in existing studies \cite{cohn2011your,cohn2012ultra,humantenna,yan2017application}.

\vspace{.5em}
\noindent {\bf Proximity requirement:}
From the measurement study in \sect\ref{sec:measurement} and the evaluation results in \sect\ref{subsec:near-field}, our approach requires that the authenticator and the authenticatee are in proximity on the same human body. For instance, in the example of personalizing smart objects, the user should use the hand with the wrist wearable to touch the objects.
A wireless reader needs to be placed close to a worn medical sensor to be authenticated. We believe that this proximity requirement introduces little overhead of using TouthAuth-based devices. Nevertheless, TouchAuth offers a low cost and small form factor solution based on ubiquitous ADCs only. Although exiting IBC and physiological sensing approaches may not have this proximity requirement, they generally require non-trivial sensing devices that are more costly and of larger form factors. In particular, the proximity requirement increases the barrier for active attackers to steal the iBEP signals, since they have to place a sensor close to the authenticator. In contrast, if the body-area property is effective for the whole body like for ECG/PPG, the attackers may attach a miniature sensor to the clothing of the victim to steal the signal.

\vspace{.5em}
\noindent {\bf iBEP injection attack:}
If an attacker can generate a strong ac EF that overrides the ambient EF, the attacker can infer the $s(t)$ sensed by the authenticator and spoof it to accept an invalid authenticatee. However, the strong EF generation is non-trivial and inevitably requires bulky equipment. Overriding the power grid voltage is generally impossible unless the building's power network is disconnected from the mains grid and supplied by a power generator controlled by the attacker. Another possible approach is to surround the victim TouchAuth devices with two metal plates connected with an ac generator. The bulky setting of the EF generation renders the attack easily discernible by the TouchAuth user and costly, unattractive to the attacker. {\rev Another possible attack is to generate power surges in the power network by frequently switching on and off high-power appliances like space heaters. However, the surges will also generate easily discernible disturbances to other appliances such as lights and audio systems. Thus, we believe that the iBEP injection attack, though possible, is unrealistic or easily discernible.}

{\rev
\vspace{.5em}
\noindent {\bf Other interferences:}
TouchAuth is based on the instantaneous similarity of the iBEP signals in close proximity induced by the ambient EF. Hence, the similarity does not depend on the user's physiological state. Certain limited scenarios may affect the iBEP. For example, a temporary charging caused by taking off a sweater may override the iBEP in a short time. However, such situations do not happen frequently.
}

\vspace{.5em}
\noindent {\bf Applicability to implantable medical devices (IMDs):}
Our measurement study (\sect\ref{sec:measurement}) and evaluation (\sect\ref{sec:eval}) are based on iBEPs collected from skins. We now discuss the applicability of TouchAuth to devices implanted into human bodies. From our discussion in \sect\ref{subsec:body-antenna}, an electrostatically induced human body is an equipotential body. Thus, the ADC pin and the ground of an IMD that is fully implanted into the human body will have the same potential. As a result, the iBEP measurement will be zero. We conducted a set of experiments to verify this. We bought two types of homogeneous meat from a supermarket. We wrapped a Z1 mote using cling film but leaved its ADC-connected electrode out of the wrap. We {\em fully} and {\em partially} implanted the mote into the meat. Under both settings, the electrode has significant contact with the meat. The partial implanting means that a small portion of the cling film was still visible.
The peak-to-peak amplitudes of the iBEP signals measured by the Z1 mote fully and partially implanted are about $0.05\,\text{V}$ and $0.1\,\text{V}$, respectively.
The peak-to-peak amplitude of the latter case is comparable to some of our measurement results on human skins (cf.~\sect\ref{sec:measurement}). Frequency analysis shows that the former is close to white noise and the latter clearly exhibits a frequency of $50\,\text{Hz}$.
These results suggest that TouchAuth is applicable to partially implanted devices, such as insulin pumps, cochlear implants, foot drop implants, etc.

\section{Related Work}
\label{sec:related}

\noindent
{\bf Device authentication and key generation:}
Various physiological signals have been exploited for contact-based {\em device authentication} and {\em key generation}.
Key generation establishes a secret symmetric key for a pair of nodes on the same human body.
Using ECG and PPG for the above two tasks has received extensive research. An early work \cite{poon2006novel} encodes the interpulse intervals (IPIs) of ECG or PPG into a bit sequence and performs authentication by comparing the Hamming distance of two bit sequences with a threshold. The study \cite{imdguard} generates IPI-based symmetric key for an IMD and an external device.
PSKA \cite{pska} and OPFKA \cite{opfka} generate keys from certain ECG/PPG features. Rostami et al. \cite{h2h} quantify ECG's randomness in terms of entropy and design the H2H authentication protocol.
However, ECG/PPG sensors often have large form factors due to the required physical distances between electrodes.
Moreover, ECG/PPG sensing can be vulnerable to video analytics \cite{poh2010non,wu2012eulerian}.
Table~\ref{tab:compare} compares the performance of APCC-TouchAuth (from Fig.~\ref{fig:eva-window-len}) and several ECG/PPG device authentication approaches.
TouchAuth achieves comparable detection accuracy within shorter sensing times. Recent studies have also exploited EMG \cite{SecretfromMuscle}
and gait \cite{walkie-talkie} for key generation. However, the multi-electrode EMG sensor \cite{SecretfromMuscle} is sizable and must be placed close to muscles.
Walking to generate keys \cite{walkie-talkie} may be inconvenient and the used inertial measurement units (IMUs) may be vulnerable to remote acoustic attack \cite{walnut}.

\begin{table}
  \caption{Comparison with existing approaches.}
  \label{tab:compare}
  \small
  \begin{tabular}{ccccc}
    \hline
    Ref. & Signal & Sensing time (s) & $\alpha$ (\%) & $\beta$ (\%) \\
    \hline
    \multicolumn{2}{c}{\bf TouchAuth} & {\bf 1} & {\bf 2.0} & {\bf 94.2\%} \\
           &  & {\bf 5} & {\bf 2.0} & {\bf 98.9\%} \\
           \hline
     \cite{poon2006novel} & ECG+PPG & $\sim$60 (67 IPIs) & 2.1 & 93.5 \\
          & & $\sim$30 (34 IPIs) & 4.5 & 90.5 \\
          \hline
\cite{pska} & PPG & 12.8 & 0.1 & 99.9 \\
          \hline
          \cite{opfka} & ECG & $\sim$90 (90 IPIs) & $\sim$0$^*$ & $\sim$100$^*$ \\
    \hline
    \end{tabular}

  $^*$ \cite{opfka} fuzzily states that its FAR and FRR are almost zero.
  \end{table}

\vspace{0.5em}
\noindent
{\bf Human body coupled capacitive sensing:} The iBEP sensing belongs to a broader area of capacitive sensing. A recent survey \cite{grosse2017finding} provides a taxonomy of capacitive sensing. We review those on passively sensing the mutual impact between the human body and ambient EF. The iBEP has been used for touch \cite{cohn2011your} and motion sensing \cite{cohn2012ultra}, keyboard stroke detection \cite{elfekey2013design}, gesture recognition \cite{humantenna}, wearables clock synchronization \cite{yan2017application}. Several studies use a single off-body electrode to sense the change of ambient EF due to human's electrophysiological signals \cite{prance2008remote} and body movement \cite{takiguchi2007human,prance2012position}. Platypus \cite{grosse2016platypus} uses an EF sensor array on the ceiling to localize and identify a human walker. The EF change is due to the triboelectric effect and changes in capacitive coupling between the walker and the environment.
{\rev Wang et al. \cite{MagnifiSense} use an external sound card as the ADC and three magneto-inductive coil sensors to collect the electromagnetic interference (EMI) radiated from various devices. The signatures contained in the EMI are used for identifying the device the user is touching. 
Laput et al. \cite{em-sense15} attach a modified software-defined radio to the human body for sampling iBEP. When the user touches an object, the class of the object can be recognized based on the sampled iBEP signal. Yang et al. \cite{em-id16} develop a follow-up research of \cite{em-sense15} to recognize the identity, rather than the class, of the touched object. However, the needed training phase of \cite{MagnifiSense,em-sense15,em-id16} introduces overhead.}

The human body can be used as a communication channel. Early studies \cite{zimmerman1995personal,matsushita2000wearable,park2006tap,baldus2009human} build customized transmitter and receiver for intra-body communication (IBC). Vu et al. \cite{vu2012distinguishing} design a wearable transmitter to convey identification data to a touchscreen as the receiver. Holz et al. \cite{holz2015biometric} use a wrist wearable and touchscreen to measure bioimpedance and identify the user. Hessar et al. \cite{hessar2016enabling} uses fingerprint scanner and touchpad as the transmitter and a software-defined radio attached to skin as the receiver. {\rev Yang et al. \cite{em-comm17} show that the transmitters can be LEDs, buttons, I/O lines, LCD screens, motors, and power supplies. Roeschlin et al. \cite{touchpairing} design an IBC approach that estimates the body channel characteristics to pair on-body devices.} Although IBC can be used for contact-based device authentication, it often requires non-trivial transmitter/receiver devices. In contrast, our approach requires a ubiquitous low-speed ADC only.

{\rev
\vspace{0.5em}
\noindent
{\bf Other related studies:} 
VAuth \cite{Feng2017} uses a wearable token device to sense the vibrations caused by the speech of the wearer and match the vibration signal with the received voice signal. The matched vibration and voice signals are further used to verify that the voice signal received by a voice assistant is really from the token wearer.
Nyemkova et al. \cite{eda18} study the distinguishability of various electronic devices based on the fluctuations of their internal EMI.
}

\section{Conclusion}
\label{sec:conclude}

This paper explained the electrostatics of iBEP with supporting measurement results. Based on the understanding, we designed TouchAuth and evaluated its same-body contact detection performance via extensive experiments under a wide range of real-world settings. Results show that TouchAuth achieves comparable detection accuracy as existing physiological sensing approaches, but within much shorter sensing times. Moreover, the uni-electrode iBEP sensor can be miniaturized. TouchAuth offers a low-cost, lightweight, and convenient approach for the authorized users to access the smart objects found in indoor environments.

{\rev
\begin{acks}
The authors wish to thank our shepherd Dr. Alanson Sample and the anonymous reviewers for providing valuable feedback on this work. This research was funded by a Start-up Grant at Nanyang Technological University.
\end{acks}
}

\clearpage

\balance
\bibliographystyle{ACM-Reference-Format}
\bibliography{001_reference} 

\end{document}